\documentclass[a4paper,11pt]{article}
\pdfoutput=1 

\usepackage{jheppub} 

\usepackage[T1]{fontenc} 

\usepackage{braket}

\newcommand{\fd}[2]{\parbox{#1}{\includegraphics[width=#1]{#2}}}

\makeatletter
\gdef\@fpheader{}
\makeatother

\title{\boldmath Hyperbolic three-string vertex}

\author{Atakan Hilmi F{\i}rat}

\affiliation{
	Center for Theoretical Physics, Massachusetts Institute of Technology,\\
	Cambridge, MA 02139, U.S.A.}

\emailAdd{firat@mit.edu}

\abstract{We begin developing tools to compute off-shell string amplitudes with the recently proposed hyperbolic string vertices of Costello and Zwiebach. Exploiting the relation between a boundary value problem for Liouville's equation and a monodromy problem for a Fuchsian equation, we construct the local coordinates around the punctures for the generalized hyperbolic three-string vertex and investigate their various limits. This vertex corresponds to the general pants diagram with three boundary geodesics of unequal lengths. We derive the conservation laws associated with such vertex and perform sample computations. We note the relevance of our construction to the calculations of the higher-order string vertices using the pants decomposition of hyperbolic Riemann surfaces.}


\arxivnumber{2102.03936}

\begin{document} 
\maketitle
\flushbottom

\section{Introduction} \label{sec:Intro}

Defining off-shell amplitudes in closed string field theory requires selecting a set of string vertices $\mathcal{V}_{g,n}$ with $2g-2+n > 0$~\cite{Zwiebach:1992ie,Erler:2019loq}. These are subsets of the moduli spaces $\widehat{\mathcal{P}}_{g,n}$ of compact Riemann surfaces of genus $g$ and $n$ punctures with a choice of local coordinates (defined up to global phases) around each puncture. String vertices ought to satisfy the geometric master equation in order to define a consistent quantum theory~\cite{Sen:1994kx,Sen:1993kb,Sonoda:1989wa}.

There have been few proposals in the past for how to explicitly specify string vertices $\mathcal{V}_{g,n}$. The oldest, and probably the most well-known, is the one that uses the minimal area metrics on Riemann surfaces~\cite{Zwiebach:1992ie,Zwiebach:1990nh}. Using such metrics there is a simple prescription for how to specify string vertices that solves the geometric master equation~\cite{Zwiebach:1992ie}. The minimal area metrics for higher genus surfaces, however, are not known explicitly and still lack rigorous proof of existence. Nonetheless, one may expect that these will soon follow in the light of the recent discoveries~\cite{Headrick:2018ncs,Headrick:2018dlw,Naseer:2019zau}.

Another proposal for string vertices $\mathcal{V}_{g,n}$ that utilizes the fact that the Riemann surfaces considered for $\mathcal{V}_{g,n}$ admit hyperbolic metrics (of constant negative Gaussian curvature $K=-1$) was recently made by Moosavian and Pius~\cite{Moosavian:2017qsp,Moosavian:2017sev}. This interesting approach seems particularly promising considering the rigorously established existence of hyperbolic metrics and the recent developments in evaluating integrals over the moduli spaces of Riemann surfaces using the associated Teichm\"uller spaces~\cite{mcshane1998simple,mirzakhani2007weil,mirzakhani2007simple,Eynard:2007fi,ellegard1,ellegard2,Dijkgraaf:2018vnm}. However, it has been shown that these string vertices solve the geometric master equation only to the first approximation and they require a correction at each order of approximation. It is not known that such corrected string vertices always exist.

Although they are intriguing in their own rights, we see that two proposals for string vertices above suffer from either missing the proof of existence or failing to satisfy the geometric master equation exactly, therefore falling short of providing a consistent string field theory. In order to have a consistent string field theory we must guarantee that the string vertices exist on the moduli spaces of Riemann surfaces while exactly satisfying the geometric master equation. \emph{Hyperbolic string vertices} by Costello and Zwiebach simultaneously achieved both of these conditions recently~\cite{Costello:2019fuh}. To that end, the authors considered Riemann surfaces endowed with hyperbolic metric with \emph{geodesic boundaries} of length $L$, for $0<L\leq2$ arcsinh(1), and with systole\footnote{Systole on a bordered surface is defined as the length of the shortest closed geodesic that is \emph{not} a boundary component.} greater than or equal to $L$. Then they specified the string vertices by attaching flat semi-infinite cylinders of circumference $L$ at each boundary component to such surfaces. By the existence of hyperbolic metrics on Riemann surfaces of genus $g$ and $n$ boundaries with $2g-2+n > 0$, it was argued that this construction is always possible. Furthermore, it has been shown that the resulting string vertices exactly satisfy the geometric master equation by the virtue of the collar theorems of hyperbolic geometry~\cite{buser2010geometry}. We are going to call the closed bosonic string field theory hyperbolic string vertices define \emph{hyperbolic string field theory}.

Beyond establishing the first rigorous, explicit, and exact construction for the string vertices, using the hyperbolic string vertices also seems promising from the perspective of the aforementioned developments in computing integrals over the moduli spaces of Riemann surfaces by exploiting the underlying hyperbolic geometry, just like in the case of the vertices of Moosavian and Pius. One might imagine (or hope) similar methods can be applied to evaluate the string amplitudes to arbitrary orders and provide a useful handle for the computations in hyperbolic string field theory as a result.

A natural first step in this direction would be to compute the off-shell three-string amplitudes using the hyperbolic three-string vertex $\mathcal{V}_{0,3}$, which is constructed by grafting three flat semi-infinite cylinders to the three-holed sphere (or \emph{pair of pants}) equipped with a hyperbolic metric, since $\mathcal{V}_{0,3}$ contains just a single surface. For the sake of generality, we are going to leave the circumferences of the grafted cylinders arbitrary for this vertex, even though only the case of equal circumferences is needed for $\mathcal{V}_{0,3}$~\cite{Costello:2019fuh}. This \emph{generalized hyperbolic three-string vertex} is of interest in the hyperbolic string field theory in the long run on account of the well-known pants decomposition of Riemann surfaces~\cite{buser2010geometry}. For brevity, we will also denote this generalized vertex as hyperbolic three-string vertex without making a distinction.

In order to perform the computations mentioned above using the operator formalism of conformal field theory (CFT), one needs to obtain the explicit expressions of the local coordinates around the punctures for the hyperbolic three-string vertex~\cite{Erler:2019loq}. In this paper, we find these local coordinates, investigate their various limits, and derive the associated conservation laws by following the procedure in~\cite{rastelli2001tachyon}. 

In principle, the local coordinates for the hyperbolic three-string vertex can be obtained by the following procedure. First, recall that the hyperbolic metric on the three-holed sphere with geodesic boundaries of lengths $L_i$ ($i=1,2,3$) is unique up to isometry~\cite{buser2010geometry}. So we can simply write down this hyperbolic metric as
\begin{equation} \label{eq:intrometric}
ds^2 = e^{\varphi(z,\bar{z})} |dz|^2,
\end{equation}
on the Riemann sphere minus three disjoint simply connected regions, or \emph{holes}, \emph{unique} up to PSL(2,$\mathbb{C}$) transformations, whose boundaries are geodesics of given lengths $L_i$. From this point of view, one can obtain the local coordinates by finding how punctured unit disks conformally map onto these simply connected regions, since a semi-infinite cylinder is conformal to a punctured unit disk and it canonically introduces the local coordinates~\cite{Costello:2019fuh}. Note that such conformal transformations exist by the Riemann mapping theorem.

Therefore, we see that the problem of finding the local coordinates for the hyperbolic three-string vertex is a two-step procedure:
\begin{enumerate}
	\item Find an explicit description of the union of three disjoint simply connected regions on the Riemann sphere whose complement is endowed with a hyperbolic metric~\eqref{eq:intrometric} and boundary components are geodesics of lengths $L_i$,
	\item Find the conformal transformations from punctured unit disks to the aforementioned simply connected regions.
\end{enumerate}
The first step clearly involves solving a complicated boundary value problem for a partial differential equation, \emph{Liouville's equation}, and getting an exact answer is a hard endeavor in general. Luckily, it is known that the solutions for such boundary value problem can be related to a monodromy problem of a particular second-order linear ordinary differential equation with regular singularities, or a \emph{Fuchsian equation}, on the complex plane~\cite{hadasz2003polyakov,hadasz2004classical}. Exploiting this relation, which we review and expand in sections~\ref{sec:Fuchsian} and~\ref{sec:Monodromy}, we find the explicit description of the hyperbolic metric~(\ref{eq:intrometric}) and of the three holes on the Riemann sphere, up to PSL(2,$\mathbb{C}$) transformations.

Furthermore, the second step becomes trivial after we find such explicit description as we argue in section~\ref{sec:Monodromy}. In the end, for the hyperbolic three-string vertex whose grafted flat cylinders have the circumferences
\begin{equation} \label{eq:circum}
L_i \equiv 2 \pi \lambda_i,
\end{equation}
we obtain the following local coordinates $w_i(z)$ around the punctures at $z=0,1,\infty$ respectively:\footnote{We fix the locations of the punctures to $z=0,1,\infty$ using PSL(2,$\mathbb{C}$) transformations without loss of generality. So indices and numbers $i,j=1,2,3$ appearing on the objects denote the punctures $z=0,1,\infty$ respectively, unless otherwise stated. This shall be obvious from the context and we are not going to report it every time. If there is no label on an object, it should be understood that it has the same value for each puncture.}
\begin{subequations} \label{eq:lc}
\begin{align} \label{eq:lc1}
w_1(z) &= \frac{1}{N_1} \exp\left( \frac{v(\lambda_1,\lambda_2, \lambda_3)}{\lambda_1} \right) z (1-z)^{-\frac{\lambda_2}{\lambda_1}} \nonumber \\
&\qquad \qquad \times \left[ \frac{_2F_1 \left(\frac{1}{2}(1+i \lambda_1 -i \lambda_2 +i \lambda_3), \, \frac{1}{2}(1+i \lambda_1 -i \lambda_2 -i \lambda_3);\, 1+ i \lambda_1;\, z\right)}{_2F_1 \left(\frac{1}{2}(1-i \lambda_1 +i \lambda_2 -i \lambda_3),\,\frac{1}{2}(1-i \lambda_1 +i \lambda_2 +i \lambda_3);\, 1- i \lambda_1;\, z\right)} \right]^{\frac{1}{i \lambda_1}} \nonumber \\
&= \frac{1}{N_1} \exp\left( \frac{v(\lambda_1,\lambda_2, \lambda_3)}{\lambda_1} \right) \left(z + \frac{1+ \lambda_{1}^2 + \lambda_{2}^2 - \lambda_{3}^2}{2(1+\lambda_{1}^2)} z^2 + \mathcal{O}(z^3)\right) ,\\
\label{eq:lc2}
\quad \;\; w_2(z) &= \frac{1}{N_2} \exp\left( \frac{v(\lambda_2,\lambda_1, \lambda_3)}{\lambda_2} \right)(1-z) z^{-\frac{\lambda_1}{\lambda_2}} \nonumber\\
&\qquad \qquad \times 
\left[ \frac{_2F_1 \left(\frac{1}{2}(1+i \lambda_2 -i \lambda_1 +i \lambda_3),\,\frac{1}{2}(1+i \lambda_2 -i \lambda_1 -i \lambda_3);\, 1+ i \lambda_2;\, 1-z\right)}{_2F_1 \left(\frac{1}{2}(1-i \lambda_2 +i \lambda_1 -i \lambda_3),\,\frac{1}{2}(1-i \lambda_2 +i \lambda_1 +i \lambda_3);\, 1- i \lambda_2;\, 1-z\right)} \right]^{\frac{1}{i \lambda_2}}\nonumber\\
&= \frac{1}{N_2} \exp\left( \frac{v(\lambda_2,\lambda_1, \lambda_3)}{\lambda_2} \right)\left((1-z) + \frac{1+ \lambda_{2}^2 + \lambda_{1}^2 - \lambda_{3}^2}{2(1+\lambda_{2}^2)} (1-z) ^2 + \mathcal{O}((1-z) ^3)\right),  \\
\label{eq:lc3}
w_3(z) &= \frac{1}{N_3} \exp\left( \frac{v(\lambda_3,\lambda_2, \lambda_1)}{\lambda_3} \right)\left(\frac{1}{z}\right) \left(1-\frac{1}{z}\right)^{-\frac{\lambda_2}{\lambda_3}} \nonumber\\
&\qquad \qquad \times 
\left[ \frac{_2F_1 \left(\frac{1}{2}(1+i \lambda_3 -i \lambda_2 +i \lambda_1),\,\frac{1}{2}(1+i \lambda_3 -i \lambda_2 -i \lambda_1);\, 1+ i \lambda_3;\, \frac{1}{z}\right)}{_2F_1 \left(\frac{1}{2}(1-i \lambda_3 +i \lambda_2 -i \lambda_1),\,\frac{1}{2}(1-i \lambda_3 +i \lambda_2 +i \lambda_1);\, 1- i \lambda_3;\, \frac{1}{z}\right)} \right]^{\frac{1}{i \lambda_3}} \nonumber \\
&= \frac{1}{N_3} \exp\left( \frac{v(\lambda_3,\lambda_2, \lambda_1)}{\lambda_3} \right) \left(\frac{1}{z} + \frac{1+ \lambda_{3}^2 + \lambda_{2}^2 - \lambda_{1}^2}{2(1+\lambda_{3}^2)}\frac{1}{z^2}  + \mathcal{O}\left(\frac{1}{z^3} \right)\right).
\end{align}
\end{subequations}
Here ${_2}F_1(a,b;c;z)$ is the ordinary hypergeometric function 
\begin{equation} \label{eq:hypergeometric}
{_2}F_1(a,b;c;z) = 1 + \frac{ab}{c} \frac{z}{1!} +  \frac{a(a+1)b(b+1)}{c(c+1)} \frac{z^2}{2!} + \cdots,
\end{equation}
and the function $v\left(\lambda_{1}, \lambda_{2}, \lambda_{3}\right) $ is given in terms of the gamma function $\Gamma(z)$ as
\begin{align}
v\left(\lambda_{1}, \lambda_{2}, \lambda_{3}\right) &\equiv \frac{1}{2i} \text{Log}\left[\frac{\Gamma\left(-i \lambda_{1}\right)^{2}}{\Gamma\left(i \lambda_{1}\right)^{2}} \frac{\gamma\left(\frac{1}{2}(1+i \lambda_{1}+i \lambda_{2}+i \lambda_{3})\right) \gamma\left(\frac{1}{2}(1+i \lambda_{1}-i \lambda_{2}+i \lambda_{3})\right)}{\gamma\left(\frac{1}{2}(1-i \lambda_{1}-i \lambda_{2}+i \lambda_{3})\right) \gamma\left(\frac{1}{2}(1-i \lambda_{1}+i \lambda_{2}+i \lambda_{3})\right)} \right],\nonumber \\
\gamma(x) &\equiv \frac{\Gamma(1-x)}{\Gamma(x)}.
\end{align}
The factors $N_i$ above will be called \emph{scale factors}. They are fixed by integer $\tilde{l}_i \in \mathbb{Z}$ via 
\begin{equation}
N_i = \exp\left[ \frac{\pi}{\lambda_i} \left(\tilde{l}_i + \frac{1}{2}\right) \right].
\end{equation}
Tilde on the integers $\tilde{l}_i$ comes from our construction. As we will see, among the sets of integers $l_i$, only specific ones, $l_i=\tilde{l}_i$, would give the correct scale factor. The integers $\tilde{l}_i$ can be determined for a given set of $\lambda_i$'s in principle. Even though we couldn't find a closed-form expression for the integers $\tilde{l}_i$'s for arbitrary $\lambda_i$'s, one can still easily find them by investigating numerical plots of the local coordinates. In certain symmetric situations it is possible to find the integers $\tilde{l}_i$'s without resorting the plots. For example, when $0 \leq \lambda_i = \lambda \leq 10$ for the grafted cylinders, we find $\tilde{l}_i=\tilde{l}=-1$. This result is anticipated to hold for all values of $\lambda_i = \lambda$.

Since we have the explicit expressions of the local coordinates for the hyperbolic three-vertex~\eqref{eq:lc}, we can check their consistency with the other local coordinates in the literature by investigating their limiting behaviors~\cite{sonoda1990covariant,Moosavian:2017qsp,Zwiebach:1988qp}. For instance, as argued in~\cite{Costello:2019fuh}, this vertex must produce the three-string vertex obtained from the minimal area metric as all $\lambda_i \to \infty$ at the same rate, which we are going to denote as the \emph{minimal area limit}. In the light of this fact, we consider the minimal area limit of the coordinates~\eqref{eq:lc} and show that the local coordinates for the minimal area three-string vertex and those for the hyperbolic three-string vertices with $\lambda \to \infty$ match perturbatively to the order $\mathcal{O}(z^{10})$ in section~\ref{sec:Limits}. We then discuss the possibility of extending our argument to all orders in $z$.

Moreover, the hyperbolic three-string vertex must reduce to the three-string vertex considered by Moosavian and Pius~\cite{Moosavian:2017qsp} as $\lambda_i \to 0$ after a suitable modification, since the geodesic boundaries become cusps in this regime and this is exactly what is considered there. We argue that this limiting behavior indeed holds in section~\ref{sec:Limits}. Lastly, we consider the situation $\lambda_{2} = \lambda_1 + \lambda_3$ with $\lambda_i \to \infty$ for which the geometry resembles the light-cone vertex~\cite{Zwiebach:1988qp}. We show that the hyperbolic three-string vertex reduces to the light-cone vertex in this limit, in accord with our expectations.

Having an explicit expression for the local coordinates~\eqref{eq:lc} also means that it is possible to derive the conservation laws for the hyperbolic three-string vertex in the spirit of~\cite{rastelli2001tachyon}, which we do in section~\ref{sec:Conservation}. Again, we can investigate various limits of these conservation laws. Especially we observe that all of our expressions in section~\ref{sec:Conservation} reduces to their respective counterparts in~\cite{rastelli2001tachyon} in the minimal area limit. This is consistent, since the open string Witten vertex and its closed string analog must generate the same conservation laws. It is known that conservation laws provide  systematic and easily implementable procedure for computations in the cubic open string field theory, especially for the level truncation~\cite{Gaiotto:2002wy}, and we hope that these expressions will accomplish the same in the hyperbolic string field theory in the future.

As a sample computation using the local coordinates~\eqref{eq:lc}, we calculate the $t^3$ term in the closed string tachyon potential $V$ with $t$ is the zero-momentum tachyonic field in the case of $\lambda_i=\lambda$. Remember this is the case that appears in the string action. We find ($\alpha'=2$)
\begin{align} \label{eq:tachyon}
&\kappa^2 V =  -t^2 + \frac{1}{3} \frac{t^3}{r^6} + \dots = -t^2 +\frac{1}{3} \exp\left[\frac{ 6v(\lambda,\lambda,\lambda) + 3\pi}{\lambda}  \right] t^3 + \cdots.
\end{align} 
Here $\kappa$ is the closed string coupling constant and $r$ is the mapping radius of the local coordinates, whose inserted expression is derived in section~\ref{sec:Monodromy}. Note that this calculation is exactly like in~\cite{kostelecky1990collective,belopolsky1995off,yang2005closed}, the only difference being the mapping radii we used for the expression above.

In order to get a sense of its value, let us set $\lambda= L_{\ast}/(2 \pi) = \text{arcsinh}(1)/\pi \approx 0.28055$, which is the largest value of $\lambda$ for which the hyperbolic vertices solves the geometric master equation~\cite{Costello:2019fuh}. Substituting this value and evaluating, we obtain the closed string tachyon potential $V$ in the hyperbolic string field theory is given by
\begin{equation} \label{eq:tp}
\kappa^2 V \approx -t^2 + \left( 1.62187 \times 10^8\right)  t^3 + \cdots.
\end{equation}
The coefficient for the $t^3$ term is quite large compared to the corresponding one in the minimal area three-string vertex, which is approximately equal to $1.602$~\cite{kostelecky1990collective,belopolsky1995off,yang2005closed}. However, this coefficient in fact has the expected order of magnitude. We can see this by considering the coefficient obtained from the minimal area three-string vertex with stubs of length $\pi$, which roughly \emph{looks like} a hyperbolic three-string vertex geometrically. The coefficient for the case with stubs is easily obtained by observing that adding stubs scales mapping radii by $e^{-\pi}$, and in turn multiplies the no-stub coefficient by $e^{6 \pi}$ by the first equality in~\eqref{eq:tachyon}. This gives approximately $e^{6\pi} \cdot 1.602  \approx 2.460 \times 10^8$, which is close to the value given in~\eqref{eq:tp}.

The outline of the paper is as follows. In section~\ref{sec:Fuchsian} we introduce the boundary value problem for the hyperbolic metric with geodesic boundaries of fixed lengths on the Riemann sphere minus three holes and its relation to Fuchsian equations. In section~\ref{sec:Monodromy} we consider the relevant monodromy problem in order to find the explicit description of the holes on the Riemann sphere. The results of these two sections are well-established in the literature~\cite{hadasz2003polyakov,hadasz2004classical}, but we provide a self-contained discussion where we emphasize and investigate the resulting hyperbolic geometry in more detail. Additionally, we construct the local coordinates around the punctures for the hyperbolic three-string vertex in section~\ref{sec:Monodromy} and later in section~\ref{sec:Limits} we investigate their various limits. Lastly, we obtain the conservation laws associated with the hyperbolic three-string vertex in section~\ref{sec:Conservation}. We conclude the paper and discuss the possible future directions in section~\ref{sec:Conc}.

\section{Liouville's equation on a three-holed sphere} \label{sec:Fuchsian}

In this section, we describe the problem of finding an explicit description of the hyperbolic metric on the three-holed sphere with geodesic boundaries of lengths $L_i$ on the Riemann sphere, which will help us obtain the shapes and locations of the geodesic boundaries and the local coordinates later on. As we mentioned briefly, this is equivalent to solving Liouville's equation with specified boundary conditions on the Riemann sphere minus three holes. This problem is hard by itself, so instead we introduce a stress-energy tensor (in the sense of Liouville theory) and consider its associated Fuchsian equation, which we define below. The properties of this equation is investigated. Most importantly, we show that its multi-valued solutions can be related to hyperbolic metrics. The results of this section are well-known in the literature in the context of Liouville theory and the uniformization problem~\cite{hadasz2003polyakov,hadasz2004classical,hadasz2006liouville,cantini2001proof,cantini2002liouville,cantini2003polyakov,zograf1988liouville,takhtajan2003hyperbolic,Seiberg:1990eb,Bilal:1987cq,Hadasz:2005gk,Teschner:2003at,hempel1988uniformization}, but we are going to provide a self-contained review that focuses on the issues relevant to us.

As noted above, our first goal is to solve \emph{Liouville's equation}
\begin{equation} \label{eq:LiouvilleEq}
\partial \bar{\partial} \varphi(z,\bar{z}) = \tfrac{1}{2} e^{\varphi(z,\bar{z}) },
\end{equation}
on the three-holed sphere $X$ whose boundaries are chosen to be geodesics of lengths $L_i$ of the metric~(\ref{eq:intrometric}). It can be easily seen that satisfying Liouville equation is equivalent to the metric~(\ref{eq:intrometric}) having constant negative curvature $K=-1$. We will call $e^{\varphi(z,\bar{z}) }$ the \emph{conformal factor} and take $\varphi(z,\bar{z}) \in \mathbb{R}$ always to define a real metric. Like we mentioned before, we can think the surface $X$ endowed with the metric~(\ref{eq:intrometric}) as the Riemann sphere $\widehat{\mathbb{C}} = \mathbb{C} \cup \{\infty\}$ with three disjoint simply connected regions taken out and this understanding will be implicit. So $(z, \bar{z})$ will denote the complex coordinates on $X \subset \widehat{\mathbb{C}}$.

Solving the boundary value problem described above directly is non-trivial and we won't attempt to do that. Instead, we are going to relate this problem to solving a more manageable linear ordinary differential equation. In order to do that, let the factor $\varphi$ denote a solution of Liouville's equation~(\ref{eq:LiouvilleEq}) and define the (holomorphic) \emph{stress-energy tensor} associated with $\varphi$ as follows~\cite{Seiberg:1990eb}:
\begin{equation} \label{eq:stress-energy}
T_{\varphi}(z) \equiv -\tfrac{1}{2}(\partial \varphi)^2 + \partial^2 \varphi = -2 e^{\frac{\varphi}{2}} \partial^2 e^{-\frac{\varphi}{2}}.
\end{equation}
Observe that we only wrote the dependence on $z$, and not on $\bar{z}$, of the stress-energy tensor since it can be shown that $T_{\varphi}$ is holomorphic, $\bar{\partial} T_{\varphi} = 0$, using Liouville's equation~\eqref{eq:LiouvilleEq}. Furthermore, the converse of this statement holds as well: If $T_{\varphi} = T_{\varphi}(z)$ is holomorphic, then the factor $\varphi$ defined by $(\ref{eq:stress-energy})$ solves the Liouville's equation. Lastly, we note that $T_{\varphi}(z)$ is the (classical) stress-energy tensor in the context of Liouville theory and it transforms under conformal transformation $z \to \tilde{z}(z)$ as follows~\cite{Seiberg:1990eb}: 
\begin{align} \label{eq:transofT}
T_{\varphi}(z) = \left(\frac{\partial \tilde{z}}{\partial z} \right)^2 \widetilde{T}_{\varphi}(\tilde{z}) + \{ \tilde{z},z \}.
\end{align}
Here tilde on the stress-energy tensor indicates that it is written in the $\tilde{z}$ coordinates and $\{\cdot, \cdot\}$ is the Schwarzian derivative: 
\begin{equation}
\{ \tilde{z},z\} \equiv
\frac{\partial^3\tilde{z}}{\partial \tilde{z}} - \frac{3}{2} \left(\frac{\partial^2 \tilde{z}}{\partial \tilde{z}} \right)^2.
\end{equation}
We can similarly define the anti-holomorphic stress-energy tensor $\overline{T_{\varphi}}$ by replacing $\partial \to \bar{\partial}$ in~\eqref{eq:stress-energy}. 

Now consider the following second-order linear ordinary differential equation constructed with the stress-energy tensor $T_{\varphi}(z)$ above~\cite{hadasz2003polyakov,hadasz2004classical}:
\begin{equation} \label{eq:Fuchsian}
\partial^2 \psi(z) + \tfrac{1}{2} T_{\varphi}(z) \psi (z) = 0.
\end{equation}
We will call this the holomorphic \emph{Fuchsian equation} associated with $T_{\varphi}(z)$. The reason for the name \emph{Fuchsian} will be justified in section~\ref{sec:Monodromy} when we show that the relevant $T_{\varphi}(z)$ contains at most double poles, so that the equation~(\ref{eq:Fuchsian}) has only regular singularities (i.e. \emph{Fuchsian}). Similarly, we can define the anti-holomorphic Fuchsian equation associated with $\overline{T_{\varphi}}(\bar{z})$. Considering~\eqref{eq:transofT}, in order to make the equation~(\ref{eq:Fuchsian}) conformal invariant, we are going to take the object $\psi(z)$ transforms as a conformal primary of dimension $(-\frac{1}{2},0)$. That is, we demand
\begin{equation} \label{eq:transofpsi}
\tilde{\psi} (\tilde{z}) = \left( \frac{\partial \tilde{z}}{\partial z}\right)^{\frac{1}{2}} \psi(z),
\end{equation}
under conformal transformation $z \to \tilde{z}(z)$.

Now suppose we have solved the Fuchsian equation and found two linearly independent, not necessarily single-valued, complex-valued solutions $\psi^{+}(z)$ and $\psi^{-}(z)$. We are going to always assume these solutions are normalized appropriately, in the sense that their Wronskian $W(\psi^{-},\psi^{+})$ is equal to one: 
\begin{equation} \label{eq:Wronskian}
W(\psi^{-},\psi^{+}) \equiv (\partial \psi^{+}) \psi^{-} -\psi^{+} (\partial \psi^{+}) = 1.
\end{equation}
Now define the ratio $A(z)$ of these solutions and observe that we have the relations
\begin{equation} \label{eq:ratio}
A(z) \equiv \frac{\psi^{+}(z)}{\psi^{-}(z)} \quad \iff \quad \psi^{+}(z) = \frac{A(z)}{\sqrt{\partial A(z)}}, \quad \psi^{-}(z) = \frac{1}{\sqrt{\partial A(z)}}.
\end{equation}
From this, we immediately see the stress-energy tensor can be written as follows:
\begin{align} \label{eq:TasSch}
T_{\varphi}(z) = - 2 \frac{\partial^2 \psi^{-}}{\psi^{-}} &= -2 (\partial A)^{\frac{1}{2}} \partial^2 (\partial A)^{-\frac{1}{2}} = (\partial A)^{\frac{1}{2}} \partial \left((\partial A)^{-\frac{3}{2}} \partial^2 A\right) \nonumber \\
&= \frac{\partial^3A(z)}{\partial A(z)} - \frac{3}{2} \left(\frac{\partial^2 A(z)}{\partial A(z)} \right)^2 \equiv \{ A(z),z\} \implies T_{\varphi}(z) = \{ A(z),z\}.
\end{align}
In general, it is highly non-trivial to find the function $A(z)$ for a given $T_{\varphi}(z)$ satisfying (\ref{eq:TasSch}) above. However, if we know the solutions to the Fuchsian equation (\ref{eq:Fuchsian}), we see that $A(z)$ is determined by~\eqref{eq:ratio} up to M\"{o}bius transformations. That is one utility of the Fuchsian equation. Moreover, given $A(z)$ satisfying $T_{\varphi}(z) = \{A(z),z\}$, we can find the normalized solutions for the Fuchsian equation from~(\ref{eq:ratio}) as well. Note that $A(z)$ is a scalar under conformal transformations as can be seen from~\eqref{eq:transofpsi} and~\eqref{eq:ratio}.

Also we can see that putting the stress-energy tensor $T_{\varphi}(z)$ in the form (\ref{eq:TasSch}) and knowing such $A(z)$ is advantageous on the account of the transformation property of the stress-energy tensor~\eqref{eq:transofT}. The relation (\ref{eq:TasSch}), combined with the transformation property of the Schwarzian derivative and the stress-energy tensor, allows us to find the explicit expression of the stress-energy tensor $T_{\varphi}$ in other coordinates. We will see the benefit of this observations in the next section.

Another utility of the Fuchsian equation (\ref{eq:Fuchsian}) can be understood as follows. We can easily see that $\psi = e^{-\frac{\varphi(z,\bar{z})}{2}}$ solves (\ref{eq:Fuchsian}) using the second equality in~\eqref{eq:stress-energy}. This solution of the Fuchsian equation is real and single-valued because the metric (\ref{eq:intrometric}) itself is real and single-valued. It is important to observe that such factor solves the Fuchsian equation, because this allows us to relate the linearly independent, normalized solutions $\psi^{\pm}(z)$ of the Fuchsian equation to the hyperbolic metric (\ref{eq:intrometric}). In other words, knowing $\psi^{\pm}(z)$ would suffice to construct the metric. 

Before we do that more precisely, we should first describe the multi-valuedness of the solutions $\psi^{\pm}(z)$. For our purposes, it is going to be sufficient to assume that the multi-valuedness of the solutions $\psi^{\pm}(z)$ are described by SL(2,$\mathbb{R}$) transformations, in the sense that when we go around any point $z=u \in \widehat{\mathbb{C}}$ by $(z-u)\to e^{2 \pi i} (z-u)$ the solutions are taken to be transforming as follows:
\begin{equation} 
\begin{bmatrix}
\psi^{+} \\ \psi^{-}
\end{bmatrix} \; \to \;
\begin{bmatrix}
a & b \\ c & d
\end{bmatrix}
\begin{bmatrix}
\psi^{+} \\ \psi^{-}
\end{bmatrix} \quad \text{where} \quad a,b,c,d \in \mathbb{R}, \quad ad-bc=1,
\end{equation}
unless otherwise stated. That is, we assume the values that the functions $\psi^{\pm}(z)$ attain at a given point are related by SL(2,$\mathbb{R}$) transformations like above. From this, it is easy to see that the solution $e^{-\frac{\varphi(z,\bar{z})}{2}}$ of~\eqref{eq:Fuchsian} is given by the following linear combination of $\psi^{+}(z)$ and $\psi^{-}(z)$:
\begin{equation} \label{eq:FtoL}
e^{-\frac{\varphi(z,\bar{z})}{2}} = C \frac{i}{2} ( \overline{\psi^{-}(z)} \psi^{+}(z) - \overline{\psi^{+}(z)} \psi^{-}(z) ),
\end{equation}
since this is the unique real linear combination of the solutions $\psi^{\pm}(z)$ that is invariant under SL(2,$\mathbb{R}$) transformations (i.e. single-valued). As usual, the bar over the solutions denotes the complex conjugation. Here $C$ is a real constant, which turns out to be $C=\pm1$, as we will show it shortly. With this, the following metric has constant negative curvature $K= -1$:
\begin{equation} \label{eq:firstmetric}
ds^2 = e^{\varphi(z,\bar{z})} |dz|^2 = \cfrac{-4 |dz|^2}{( \overline{\psi^{-}(z)} \psi^{+}(z) - \overline{\psi^{+}(z)} \psi^{-}(z) )^2}.
\end{equation}
Note that a version of these expressions appears in the context of Liouville theory~\cite{Seiberg:1990eb}. There, the solutions $\psi^{\pm}(z)$ are interpreted as spin-$1/2$ representations of SL(2,$\mathbb{R}$) and their physical meaning is discussed.

The main takeaway from the discussion in the previous paragraphs is that the hyperbolic metric on a three-holed sphere $X$ can be related to the solutions of the Fuchsian equation using a suitable $T_{\varphi}(z)$. From the expression in (\ref{eq:stress-energy}), it might seem that finding $T_{\varphi}(z)$ as a function of $z$ is as hard as finding the explicit form of the metric (\ref{eq:intrometric}). However, as we will see in section \ref{sec:Monodromy}, $T_{\varphi}(z)$ can be found without knowing the metric. Then we can deduce the form of the hyperbolic metric by solving the associated Fuchsian equation through the relation~\eqref{eq:firstmetric}, which will eventually lead us to the local coordinates.\footnote{These relations hold for other hyperbolic Riemann surfaces with geodesic boundaries as well. But we will restrict our discussion to three-holed sphere, since it is the simplest case to perform these computations explicitly.}

Before we conclude this section, we need to show $C = \pm 1$ as we claimed. It is clear that not every value of a priori unfixed $C \in \mathbb{R}$ can define a hyperbolic metric with $K=-1$, so we need to choose the right value(s). This is essentially the reflection of the fact that the Fuchsian equation is linear: Every scaling of $e^{-\frac{\varphi(z,\bar{z})}{2}}$ is also a solution of (\ref{eq:Fuchsian}), even though the scaled ones don't define a hyperbolic metric with $K=-1$ because the Liouville's equation~(\ref{eq:LiouvilleEq}) is non-linear.

We can fix such $C$ once and for all as follows. First note that the conformal factor
\begin{equation} \label{eq:conformalfactor}
e^{\varphi(z,\bar{z})} = \frac{\lambda^2 |\partial f (z)|^2}{|f(z)|^2 \sin^2(\lambda \log|f(z)|)} = \frac{|\partial(\lambda \log(f(z)))|^2}{\sin^2(\lambda \log|f(z)|)}.
\end{equation}
always defines a (possibly singular) hyperbolic metric with $K=-1$, or equivalently, $\varphi$ above solves the Liouville's equation (\ref{eq:LiouvilleEq}) for an arbitrary holomorphic function $f(z)$ and an arbitrary $\lambda \in \mathbb{R}_{\geq 0}$, as one can check by explicit calculation. Now take the function $f(z)$ to be equal to
\begin{equation} \label{eq:HolChoice}
f(z) = A(z)^{\frac{1}{i \lambda}} =  \left( \frac{\psi^{+}(z)}{\psi^{-}(z)} \right)^{\frac{1}{i \lambda}}.
\end{equation}
We will denote the right-hand side as the \emph{scaled ratio}. After substituting this expression into~(\ref{eq:conformalfactor}) we exactly get the metric~(\ref{eq:firstmetric}). This shows $C =\pm1$. For us, the equivalence between~(\ref{eq:firstmetric}) and (\ref{eq:conformalfactor}), with the choice (\ref{eq:HolChoice}), is going to be extremely useful and we will use both forms interchangeably in our arguments.

In summary, we have seen that we can relate the hyperbolic metric on a three-holed sphere $X$ to the solutions of the Fuchsian equation (\ref{eq:Fuchsian}) through (\ref{eq:firstmetric}). Not only this will provide us a solution to the Liouville's equation (\ref{eq:LiouvilleEq}), but, more importantly, it will be also used to make the boundaries of $X$ geodesics of the metric~\eqref{eq:intrometric}. After all, that's the whole reason we are taking this detour into Fuchsian equations. We have already seen that the conformal factor~(\ref{eq:conformalfactor}) always defines a (possibly singular) hyperbolic metric for any given $f(z)$, but the boundaries of $X$ are going to be geodesics only when we relate it to a particular set of solutions for the Fuchsian equation through the relation (\ref{eq:HolChoice}), as we shall see. In the next section, we are going to focus on the three-punctured sphere $\widetilde{X} = \mathbb{C} \setminus \{0,1,\infty\}$, rather than a three-holed sphere $X$, since it is simpler to deal with initially. Then we will cut open appropriate holes around the punctures in $\widetilde{X}$ to return back to $X \subset \widehat{\mathbb{C}}$ and graft flat semi-infinite cylinders to these holes to construct the local coordinates for the hyperbolic three-string vertex.

\section{A monodromy problem of Fuchsian equation} \label{sec:Monodromy}

In this section we find the hyperbolic metric on a three-holed sphere $X$ by investigating a certain monodromy problem of the Fuchsian equation (\ref{eq:Fuchsian}) on the three punctured sphere $\widetilde{X}$ and construct the local coordinates for the hyperbolic three-string vertex. First, we describe the relevant monodromy problem and solve the Fuchsian equation on $\widetilde{X}$ accordingly. Then we find the explicit form of the (singular) hyperbolic metric on $\widetilde{X}$ by the relations given in section~\ref{sec:Fuchsian}. The resulting geometry looks like three semi-infinite series of hyperbolic cylinders, attached where they flare up, connected to each other while keeping the curvature constant and negative.

Next, we cut these hyperbolic cylinders out from the geometry appropriately, which leave us with a three-holed sphere $X$. This procedure doesn't change the hyperbolic metric, so at the end we obtain an explicit description of the hyperbolic metric with geodesic boundaries on a three-holed sphere. Moreover, we describe the holes on the Riemann sphere explicitly by investigating the simple closed geodesics of this hyperbolic metric. After that, grafting flat semi-infinite cylinders needed for the construction of the local coordinates amounts to simple conformal transformations of the punctured unit disks to these holes.

Most of the results from this section (except for subsection \ref{sec:Local}) are from~\cite{hadasz2003polyakov,hadasz2004classical}, for which we provide a detailed summary. However, we elaborate the geometric picture coming from the hyperbolic metric in more detail and prove some important results necessary for the explicit construction of the local coordinates.

\subsection{Description of the monodromy problem}

Consider the three-punctured sphere $\widetilde{X} = \mathbb{C} \setminus \{0,1,\infty\}$ and suppose that the solutions of the Fuchsian equation (\ref{eq:Fuchsian}) have hyperbolic SL(2,$\mathbb{R}$) monodromy around each puncture. That is, as we go around a puncture by $(z-z_j) \to e^{2 \pi i }(z-z_j)$, we demand that the solutions for the Fuchsian equation $\psi^{\pm}(z)$ change as,
\begin{equation} \label{eq:SL2R}
\begin{bmatrix}
\psi^{+} \\ \psi^{-}
\end{bmatrix} \; \to \;
M^j
\begin{bmatrix}
\psi^{+} \\ \psi^{-}
\end{bmatrix} \quad \text{where} \quad M^j \in \text{SL(2,}\mathbb{R}), \quad |\text{Tr} M^j|>2.
\end{equation}
Note that the condition on the trace makes the matrix $M^j$ a hyperbolic element of $\text{SL(2,}\mathbb{R})$ and that's why we say we have a hyperbolic monodromies around the puncture $z=z_j$. Realizing this structure for the solutions to the Fuchsian equation and finding them is our \emph{monodromy problem}. This problem is first considered in~\cite{hadasz2004classical} in the context of Liouville theory. We will call a puncture \emph{hyperbolic singularity} if the solutions of the Fuchsian equation have a hyperbolic SL(2,$\mathbb{R}$) monodromy around it.

In order to solve the monodromy problem, we need to first determine appropriate $T_{\varphi}(z)$ as a function of $z$ (if exists) so that the solutions of the Fuchsian equation can realize these monodromies around the punctures. Then declaring that particular $T_{\varphi}(z)$ to be equal to (\ref{eq:stress-energy}) coming from Liouville theory and using the reasoning in section \ref{sec:Fuchsian} we can extract the possibly singular hyperbolic conformal factor on $\widetilde{X}$ with the solutions that realize these monodromies. As explained above the equation ~\eqref{eq:FtoL}, this metric is going to be single-valued by $\text{SL(2,}\mathbb{R})$ monodromies and it will eventually lead us to the hyperbolic metric with geodesic boundaries on a three-holed sphere $X$. 

Before we do that, let us investigate an individual hyperbolic singularity. We begin by picking a puncture, say $z=0$, and choosing a normalized basis of solution $\psi_{1}^{\pm}(z)$ for which the monodromy around $z=0$ is diagonal as follows:
\begin{equation} \label{eq:DiagMon}
\begin{bmatrix}
\psi_1^{+} \\ \psi_1^{-}
\end{bmatrix} \; \to \;
\begin{bmatrix}
-e^{-\pi \lambda_1} & 0 \\ 0 & -e^{\pi \lambda_1} 
\end{bmatrix}
\begin{bmatrix}
\psi_1^{+} \\ \psi_1^{-}
\end{bmatrix} 
\iff
\psi_{1}^{\pm}(z) = \frac{e^{\pm \frac{i v_1}{2}}}{\sqrt{i \lambda_1}} z^{\frac{1\pm i \lambda_1}{2}} (1+ \mathcal{O}(z)).
\end{equation}
The solutions always can be put into this form around $z=0$ since hyperbolic elements of SL(2,$\mathbb{R}$) can be diagonalized by conjugation, which amounts to performing a SL(2,$\mathbb{R}$) change of basis of the solutions. Here $\lambda_1 \in \mathbb{R}$ will be called the \emph{geodesic radius} associated with the $z=0$ puncture and the reason for its name will be apparent shortly. Without loss of generality we will take $\lambda_1>0$. Note that the Wronskian of these solutions is equal to $1$ thanks to the factor $1/\sqrt{i \lambda_1}$ in front. Furthermore, we also included the factors $e^{\pm \frac{i v_1}{2}}$, with $v_1 \in \mathbb{C}$, to account for the multiplicative constant that is not fixed by the Wronskian condition~\eqref{eq:Wronskian}. As we shall see, the constant $v_1$ will be fixed below by demanding SL(2,$\mathbb{R}$) monodromies around each puncture.

Using~\eqref{eq:DiagMon}, we can write the scaled ratio associated with the puncture $z=0$, as in~\eqref{eq:HolChoice},
\begin{equation} \label{eq:ScaledRatio}
\rho_1(z) \equiv \left( \frac{\psi_1^+(z)}{\psi_1^-(z)} \right)^{\frac{1}{i \lambda_1}} = e^{\frac{v_1}{\lambda_1}}(z + \mathcal{O}(z^2)).
\end{equation}
Note that this series expansion converges only on the open unit disk $D_1 = \{z \in \mathbb{C} \; | \; |z|<1 \}$, around the puncture $z=0$, since outside $D_1$ the scaled ratio $\rho_1(z)$ is multi-valued by the solutions $\psi_{1}^{\pm}(z)$ having a non-diagonal monodromy around the punctures at $z=1, \infty$. We can analytically continue the scaled ratio defined above outside the disk $D_1$, but inevitably this will require us to choose a branch for which $\rho_1(z)$ is continuous across $\partial D_1$ except at the punctures/branch cuts. We will choose the branch cut $\widetilde{L}_1$ of $\rho_1(z)$ to extend from 1 to $\infty$ along the real axis and take this to be the principal branch of $\rho_1(z)$. Thus, we conclude that the scaled ratio $\rho_1(z)$ can be defined analytically on the set
\begin{equation} \label{eq:s1}
S_1 = \mathbb{C} \setminus \widetilde{L}_1,
\end{equation}
with the expansion (\ref{eq:ScaledRatio}). When we mention the scaled ratio, we will consider the principal branch implicitly henceforth, unless otherwise stated. Lastly, note that the scaled ratio is an analytic scalar under conformal transformations, just like the ratio $A(z)$ in~\eqref{eq:ratio}.

Now by performing the conformal transformation $z \to \rho_1 = \rho_1(z)$ on $S_1$ and using the equation (\ref{eq:TasSch}) along with the properties of the Schwarzian derivative we see
\begin{equation} \label{eq:Tsect3}
T_{\varphi}(z) =\{ A(z), z\}= \left\{ \frac{\psi_1^+(z)}{\psi_1^-(z)}, z\right\} =\{\rho_1(z)^{i \lambda}, z \} = (\partial \rho_1)^2 \{ \rho_1^{i \lambda_1}, \rho_1\}+ \{\rho_1,z \}.
\end{equation}
Comparing the final form with~\eqref{eq:transofT} we read that the stress-energy tensor $\widetilde{T_{\varphi}}(\rho_1)$ in the $\rho_1$-plane takes the following form:
\begin{equation} \label{eq:closetopuncture}
\widetilde{T_{\varphi}}(\rho_1) = \{ \rho_1^{i \lambda_1}, \rho_1\} = \frac{\Delta_1}{\rho_1^2} \quad \text{where} \quad \Delta_1 \equiv \frac{1}{2} + \frac{\lambda_1^2}{2}.
\end{equation}
Here the real number $\Delta_1$ will be called the \emph{weight}. As a result of this, the Fuchsian equation in the $\rho_1$-plane takes a very simple form and we can easily obtain its solutions:
\begin{equation}
\frac{\partial^2 \tilde{\psi}(\rho_1)}{\partial \rho_1^2} + \frac{\Delta_1}{2 \rho_1^2}  \tilde{\psi}(\rho_1) =0
\implies  \widetilde{\psi}_1^{\pm} (\rho_1)= \frac{\rho_1^{\frac{1 \pm i \lambda_1}{2}}}{\sqrt{i \lambda_1}}.
\end{equation}
Here, $\widetilde{\psi}_1^{\pm}(z)$ are normalized solutions that are chosen to have diagonal monodromy around the puncture $z=0$, or equivalently $\rho_1 = 0$. Here we set the phase factor not fixed by Wronskian equal to one for convenience.\footnote{Considering this factor just adds a phase shift for the sine that appears in~\eqref{eq:met}, which would be unimportant for our considerations in this subsection.} Note that the scaled ratio of these two solutions is simply
\begin{equation}
\left( \frac{ \tilde{\psi}^{+}_1(z)}{ \tilde{\psi}^{-}_1(z)} \right)^{\frac{1}{i\lambda_1}} = \rho_1.
\end{equation}
As a result, the hyperbolic metric that the Fuchsian equation produces in the $\rho_1$-and $z$-plane are simply given by, using the relation (\ref{eq:conformalfactor}) with the choice $f(\rho_1)=\rho_1$,
\begin{equation} \label{eq:met}
ds^2 =  \frac{\lambda_{1}^2 }{|\rho_1|^2 \sin^2(\lambda_1 \log|\rho_1|)} |d\rho_1|^2 = \frac{\lambda_{1}^2 |\partial \rho_1 (z)|^2}{|\rho_1(z)|^2 \sin^2(\lambda_1 \log|\rho_1(z)|)} |dz|^2.
\end{equation}
There are two important things we should notice here. First, the metric takes the form of a series of hyperbolic cylinders that are attached to each other where they flare up in the $\rho_1$-plane, and by the expansion (\ref{eq:ScaledRatio}), when we are sufficiently close to $z=0$ in the $z$-plane. We will explain this fact, along with the closed geodesics/singularities of this metric in more detail after we obtain the explicit form for $\rho_1(z)$.

Secondly, the $z$-plane metric is smooth (except for the singularities) not only over $S_1$ but across the branch cut $\widetilde{L}_1$ as well. The reason is simply that we demanded SL(2,$\mathbb{R}$) monodromy around each puncture and we know that the metric above is invariant under the monodromies of that kind by the equivalent form in (\ref{eq:firstmetric}). So we can use the metric above in the entirety of the $z$-plane minus punctures as long as we guarantee the SL(2,$\mathbb{R}$) monodromies around all punctures simultaneously.

Since we are also demanding hyperbolic SL(2,$\mathbb{R}$) monodromies for the remaining punctures, two facts above hold for them without too much modification. We just have to change $\rho_1$ with appropriate $\rho_j$. Moreover, these produce the same hyperbolic metric when we pullback them to the $z$-plane from any $\rho_j$-plane. This can be easily seen by noticing the fact that the appropriate SL(2,$\mathbb{R}$) change of basis of solutions $\psi_1^{\pm}(z)$ can diagonalize the monodromy around another puncture, by the fact that hyperbolic elements in SL(2,$\mathbb{R}$) are conjugate to a diagonal matrix. Such transformations of the solutions don't affect the metric as we argued before.

\begin{figure}[!t]
	\centering
	\fd{6cm}{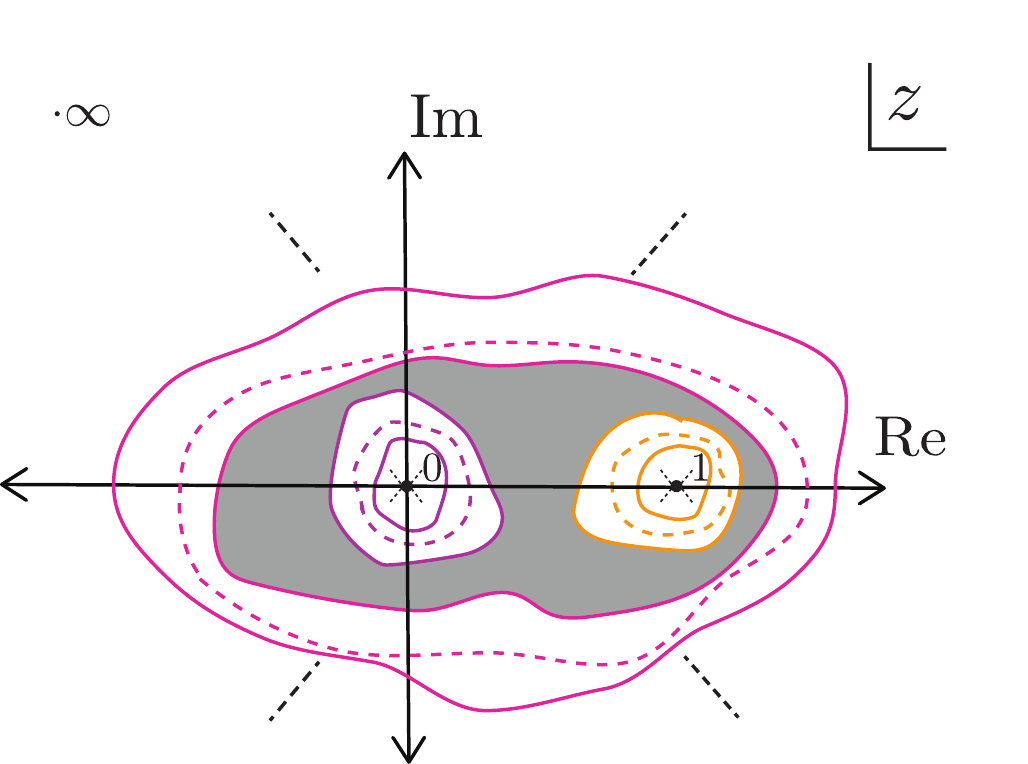}
	\fd{9cm}{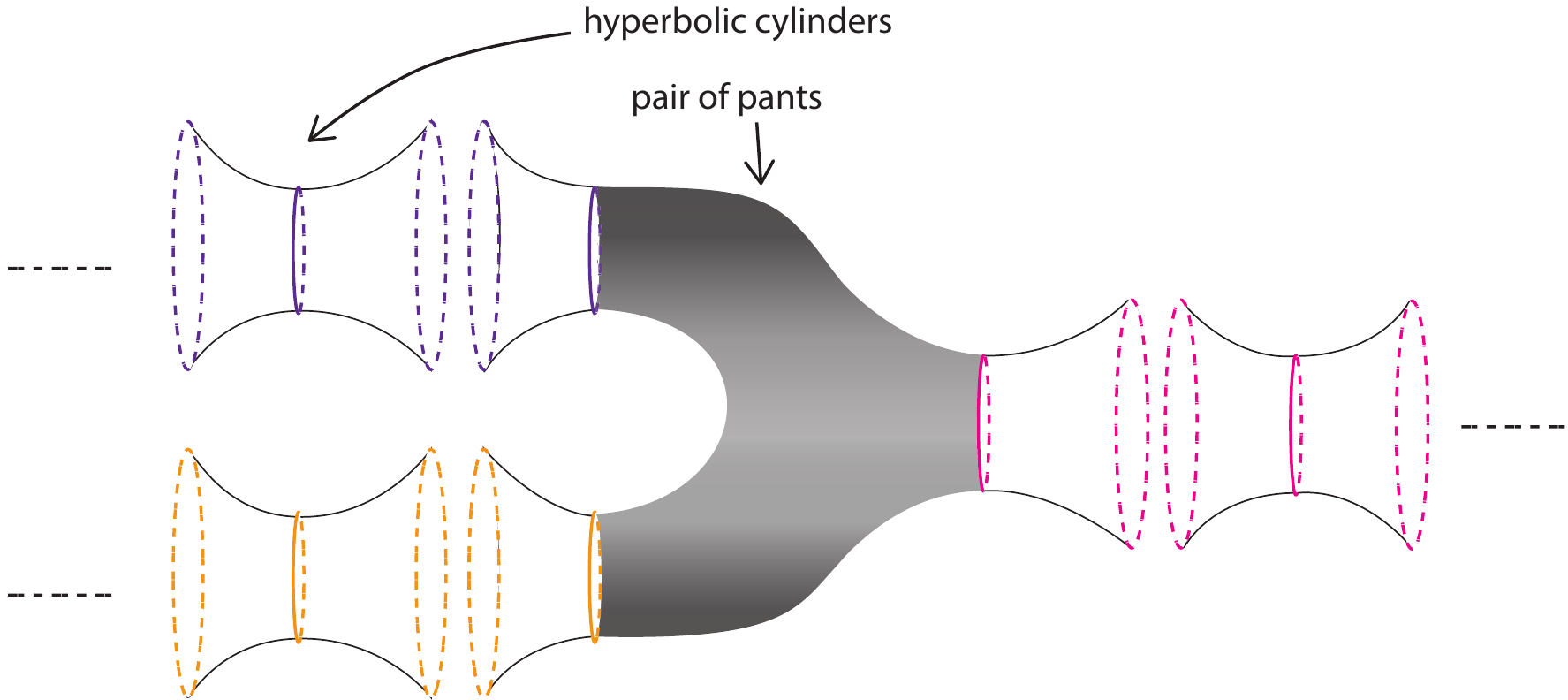}
	\caption{Sketch of the hyperbolic metric described by the Fuchsian equation with three hyperbolic singularities. The smooth hyperbolic pair of pants with geodesic boundaries is going to connect the hyperbolic cylinders, as we shall see more explicitly.}
	\label{fig:CylinderSketch}
\end{figure}

In conclusion, we see the motivation behind using the Fuchsian equation with correct monodromy structure in more detail from these comments. Even though any choice of holomorphic function works in (\ref{eq:conformalfactor}) to define a hyperbolic metric, using the scaled ratio coming from the Fuchsian equation with the monodromy data above will guarantee to generate the hyperbolic metric (\ref{eq:met}) on the $z$-plane where three series of attached hyperbolic cylinders connected to each other with hyperbolic pair of pants (i.e. three-holed sphere endowed with a hyperbolic metric), as shown in figure \ref{fig:CylinderSketch}. Moreover, it is easy to see from figure \ref{fig:CylinderSketch} that one can obtain a description of the hyperbolic pair of pants by taking out the hyperbolic cylinders and considering the remaining connected region only. This justifies why we considered this particular monodromy problem of Fuchsian equation on $\widetilde{X}$: It is a natural starting point to generate the hyperbolic metric with geodesic boundaries on $X$.

\subsection{Solution to the monodromy problem}

Before we describe the hyperbolic pair of pants, we are going to get an explicit expression for the hyperbolic metric on the three-punctured sphere $\widetilde{X}$ resulting from three hyperbolic singularities. First, we solve the monodromy problem. That is, we find $T_{\varphi}(z)$ for which the solutions of the Fuchsian equation can realize the monodromy structure described in~\eqref{eq:SL2R}. Then we solve the resulting Fuchsian equation with these prescribed monodromies and proceed to construct the metric by finding the scaled ratio.

In order to find $T_{\varphi}(z)$ as a function of $z$, observe that when we are close to the puncture $z=0$, i.e. $\rho_1 = 0$, the stress-energy tensor in~\eqref{eq:Tsect3} takes the form
\begin{equation}
T_{\varphi}(z) = (\partial \rho_1)^2 \frac{\Delta_1}{\rho_1^2} + \{\rho_1,z \} = (e^{\frac{v_1}{\lambda_1}} + \dots)^2 \frac{\Delta_1}{(e^{\frac{v_1}{\lambda_1}} z + \dots)^2} + \{e^{\frac{v_1}{\lambda_1}} z + \dots,z \} 
= \frac{\Delta_1}{z^2} + \mathcal{O}(\frac{1}{z}),
\end{equation}
using (\ref{eq:ScaledRatio}) and (\ref{eq:closetopuncture}). From this, we see that $T_{\varphi}(z)$ must have at most double poles of residues $\Delta_1,\Delta_2$ and $\Delta_3$ at $z=0,1,\infty,$ respectively in order to have a hyperbolic singularity. One can easily show that the unique $T_{\varphi}(z)$ that has such structure is
\begin{equation} \label{eq:3T}
T_{\varphi}(z) = \frac{\Delta_1}{z^2} + \frac{\Delta_2}{(z-1)^2} + \frac{\Delta_3 - \Delta_1 - \Delta_2}{z(z-1)},
\end{equation}
with $\Delta_i = (1+\lambda_i^2)/2$. Clearly we have at most double poles at $z=0,1$ with appropriate residues. Using the inversion map $z \to \tilde{z} =1/z$, along with $\{\tilde{z},z\}=0$, we can easily see that we have the correct structure at infinity, i.e. $\tilde{z} = 0$, as well:
\begin{equation}
\widetilde{T_{\varphi}}(\tilde{z}) = \frac{1}{\tilde{z}^4} \left[\Delta_1 \tilde{z}^2 + \Delta_2\tilde{z}^2 + (\Delta_3 - \Delta_1 - \Delta_2)\tilde{z}^2  + \mathcal{O}(\tilde{z}^3) \right] 
= \frac{\Delta_3}{\tilde{z}^2} +\mathcal{O}(\frac{1}{\tilde{z}}).
\end{equation}

The stress-energy tensor $T_{\varphi}(z)$ in (\ref{eq:3T}) solves the monodromy problem. In order to see that, first observe the Fuchsian equation in this case takes the form
\begin{equation} \label{eq:Fuchplugged}
\partial^2 \psi(z) + \frac{1}{2} \left[\frac{\Delta_1}{z^2} + \frac{\Delta_2}{(z-1)^2} + \frac{\Delta_3 - \Delta_1 - \Delta_2}{z(z-1)}\right] \psi (z) = 0.
\end{equation}
This is the hypergeometric equation, written in the so-called $Q$-form. The solutions of this equation and their properties are well tabulated (see Schwarz's function in \cite{hypergeometric}, also \cite{hadasz2004classical,Bilal:1987cq}). They are, with proper normalization and assignment of diagonal monodromy around $z=0$,
\begin{subequations}
\begin{align}
\psi_1^{\pm}(z) &= \frac{e^{\pm \frac{i v(\lambda_1, \lambda_2,\lambda_3)}{2}}}{\sqrt{i \lambda_1}} z^{\frac{1 \pm i \lambda_1}{2}} (1-z)^{\frac{1 \mp i \lambda_2}{2}} \nonumber \\ &\quad \times {_2}F_1 \left(\frac{1 \pm i \lambda_1 \mp i \lambda_2 \pm i \lambda_3}{2},\frac{1 \pm i \lambda_1 \mp i \lambda_2 \mp i \lambda_3}{2}; 1 \pm i \lambda_1; z\right).
\end{align}
Here ${_2}F_1(a,b;c;z)$ is the ordinary hypergeometric function~\eqref{eq:hypergeometric}. Using the transformation properties of these solutions (\ref{eq:transofpsi}) and appropriately exchanging punctures, we can also find the normalized solutions having a diagonal monodromy around $z=1$ and $z=\infty$. They are, respectively,
\begin{align}
\psi_2^{\pm}(z) &= i\frac{ e^{\pm \frac{i v(\lambda_2, \lambda_1,\lambda_3)}{2}}}{\sqrt{i \lambda_2}} (1-z)^{\frac{1 \pm i \lambda_2}{2}} z^{\frac{1 \mp i \lambda_1}{2}} \nonumber  \\ &\quad\times {_2}F_1 \left(\frac{1 \pm i \lambda_2 \mp i \lambda_1 \pm i \lambda_3}{2},\frac{1 \pm i \lambda_2 \mp i \lambda_1 \mp i \lambda_3}{2}; 1 \pm i \lambda_2; 1-z\right), \\
\psi_3^{\pm}(z) &= (iz)\frac{e^{\pm \frac{i v(\lambda_3, \lambda_2,\lambda_1)}{2}}}{\sqrt{i \lambda_3}} \left(\frac{1}{z}\right)^{\frac{1 \pm i \lambda_3}{2}} \left(1-\frac{1}{z}\right)^{\frac{1 \mp i \lambda_2}{2}} \nonumber\\ &\quad \times {_2}F_1 \left(\frac{1 \pm i \lambda_3 \mp i \lambda_2 \pm i \lambda_1}{2},\frac{1 \pm i \lambda_3 \mp i \lambda_2 \mp i \lambda_1}{2}; 1 \pm i \lambda_3; \frac{1}{z}\right).
\end{align}
\end{subequations}
We should emphasize again that the constant $v(\lambda_1,\lambda_2, \lambda_3) = v_1$ above is not fixed by the Wronskian and we will determine it below by demanding hyperbolic SL(2,$\mathbb{R}$) monodromies around all punctures. We will call this \emph{compatibility} of monodromies. Notice that compatibility is not guaranteed a priori. This is because when we demand a SL(2,$\mathbb{R}$) monodromy around a puncture, the monodromies around remaining punctures are elements of SL(2,$\mathbb{C}$), rather than SL(2,$\mathbb{R}$), in general.\footnote{It can still have unit determinant without loss of generality if one assumes appropriately normalized solutions in the sense of~\eqref{eq:Wronskian}.} So, actually, in order to solve the monodromy problem completely, we must show that the compatibility is achievable for the Fuchsian equation~\eqref{eq:Fuchplugged}.

In order to ensure compatibility, first observe that we have some SL(2,$\mathbb{C}$) monodromy around $z=1$ if we use the basis $\psi_1^{\pm}(z)$. That is, as $(1-z) \to e^{2 \pi i}(1-z)$, we have
\begin{equation}
\begin{bmatrix}
\psi_1^{+} \\ \psi_1^{-}
\end{bmatrix} \; \to \;
M_1^2
\begin{bmatrix}
\psi_1^{+} \\ \psi_1^{-}
\end{bmatrix} \quad \text{where} \quad M_1^2 \in SL(2,\mathbb{C}).
\end{equation}
Here, and throughout, we are going to denote the monodromy of the solutions $\psi_i^{\pm}(z)$ around the puncture $z=z_j$ as $M_i^j$. In order to have hyperbolic $\text{SL(2,}\mathbb{R})$ monodromy around $z=1$ while simultaneously having hyperbolic $\text{SL(2,}\mathbb{R})$ monodromy around $z=0$, we have to make sure that $M_1^2 \in \text{SL(2,}\mathbb{R})$ and $|\text{Tr} M_1^2|>2$ by adjusting $v_1$ appropriately. To that end, first observe that we have a diagonal hyperbolic $\text{SL(2,}\mathbb{R})$ monodromy around $z=1$ if we use the basis $\psi_2^{\pm}(z)$:
\begin{equation}
\begin{bmatrix}
\psi_2^{+} \\ \psi_2^{-}
\end{bmatrix} \; \to \;
M_2^2
\begin{bmatrix}
\psi_2^{+} \\ \psi_2^{-}
\end{bmatrix}
=
\begin{bmatrix}
-e^{-\pi \lambda_2} & 0 \\ 0 & -e^{\pi \lambda_2} 
\end{bmatrix}
\begin{bmatrix}
\psi_2^{+} \\ \psi_2^{-}
\end{bmatrix}.
\end{equation} 
Secondly, notice that two basis $\psi_1^{\pm}(z)$ and $\psi_2^{\pm}(z)$ are related via the connection formulas for the hypergeometric function~(see section 2.9 in~\cite{hypergeometric}, also~\cite{hadasz2004classical,Bilal:1987cq})
\begin{equation} \label{eq:connection}
\begin{bmatrix}
\psi_1^{+} \\ \psi_1^{-}
\end{bmatrix} 
=
S
\begin{bmatrix}
\psi_2^{+} \\ \psi_2^{-}
\end{bmatrix}
=
\sqrt{\lambda_1 \lambda_2} \begin{bmatrix}
e^{i \frac{v_1-v_2}{2}} \; g_{-}& e^{i \frac{v_1+v_2}{2}} \; g_{+} \\ -e^{-i \frac{v_1+v_2}{2}} \; \overline{g_{+}}& -e^{-i \frac{v_1-v_2}{2}} \; \overline{g_{-}}
\end{bmatrix}
\begin{bmatrix}
\psi_2^{+} \\ \psi_2^{-}
\end{bmatrix},
\end{equation} 
here $v_2 = v(\lambda_2, \lambda_1,\lambda_3)$ and the functions $g_{\pm}$ are given by
\begin{equation} \label{eq:gpm}
g_{\pm}=\frac{\Gamma\left(i \lambda_{1}\right) \Gamma\left(\pm i \lambda_{2}\right)}{\Gamma\left(\frac{1+i \lambda_{1} \pm i \lambda_{2}+i \lambda_{3}}{2}\right) \Gamma\left(\frac{1+i \lambda_{1} \pm i \lambda_{2}-i \lambda_{3}}{2}\right)}.
\end{equation}
Using them, we observe the monodromies in two basis are related by the following conjugation:
\begin{equation} \label{eq:conj}
\begin{aligned}
M_1^2 =& S M_2^2 S^{-1} =\lambda_{1} \lambda_{2} \left[\begin{array}{cc}
\mathrm{e}^{-\pi \lambda_{2}}\left|g_{-}\right|^{2}-\mathrm{e}^{\pi \lambda_{2}}\left|g_{+}\right|^{2} & -\left(\mathrm{e}^{\pi \lambda_{2}}-\mathrm{e}^{-\pi \lambda_{2}}\right) \mathrm{e}^{i v_{1}} g_{+} g_{-} \\[10pt]
\left(\mathrm{e}^{\pi \lambda_{2}}-\mathrm{e}^{-\pi \lambda_{2}}\right) \mathrm{e}^{-i v_{1}} \overline{g_{+} g_{-}} & \mathrm{e}^{\pi \lambda_{2}}\left|g_{-}\right|^{2}-\mathrm{e}^{-\pi \lambda_{2}}\left|g_{+}\right|^{2}
\end{array}\right].
\end{aligned}
\end{equation}
Here we used the fact
\begin{equation} \label{eq:identity}
|g_+|^2 - |g_-|^2 = \frac{1}{\lambda_1 \lambda_2},
\end{equation}
which can be derived from the expression~\eqref{eq:gpm}.

Now it is a simple calculation using~\eqref{eq:conj} and~\eqref{eq:identity} to check that $\det M_1^2 =1$. Therefore in order to have $M_1^2\in$ SL(2$, \mathbb{R}$) it is enough to make sure the entries of $M_1^2$ are real. That means we have
\begin{align} \label{eq:v}
\mathrm{e}^{2 i v\left(\lambda_{1}, \lambda_{2}, \lambda_{3}\right)} =\frac{\overline{g_{+} g_{-}}}{g_{+} g_{-}}  = \frac{\Gamma\left(-i \lambda_{1}\right)^{2}}{\Gamma \left(i \lambda_{1}\right)^{2}} \frac{\gamma\left(\frac{1+i \lambda_{1}+i \lambda_{2}+i \lambda_{3}}{2}\right) \gamma\left(\frac{1+i \lambda_{1}-i \lambda_{2}+i \lambda_{3}}{2}\right)}{\gamma\left(\frac{1-i \lambda_{1}-i \lambda_{2}+i \lambda_{3}}{2}\right) \gamma\left(\frac{1-i \lambda_{1}+i \lambda_{2}+i \lambda_{3}}{2}\right)} ,
\end{align}
with the function $\gamma(x)$ defined as
\begin{equation}
\gamma(x) \equiv \frac{\Gamma(x)}{\Gamma(1-x)}.
\end{equation}
The equality~\eqref{eq:v} fixes the exponent $v(\lambda_1,\lambda_2, \lambda_{3})=v_1$, but in a rather complicated way, and shows that it is real. Moreover, we can also easily observe Tr$M_1^2 = -2 \cosh(\pi \lambda_2)$ using~\eqref{eq:conj} and~\eqref{eq:identity}, which unsurprisingly shows the monodromy is still hyperbolic. Thus, we conclude that we can have hyperbolic SL(2$, \mathbb{R}$) monodromy around $z=1$ while having a hyperbolic SL(2$, \mathbb{R}$) monodromy around $z=0$. 

Note that that guaranteeing a hyperbolic SL(2$, \mathbb{R}$) monodromies around $z=0,1$ simultaneously with the correct choice of $v(\lambda_1,\lambda_2, \lambda_{3})$ would be sufficient for guaranteeing a hyperbolic SL(2$, \mathbb{R}$) monodromy around $z=\infty$ as well, which is the only remaining point where we have a nontrivial monodromy around. This is because we can imagine a contour that surrounds both $z=0$ and $z=1$ whose associated monodromy would be a product of two hyperbolic SL(2$, \mathbb{R}$) matrices, which is another SL(2$, \mathbb{R}$) matrix. Furthermore, this monodromy would be clearly hyperbolic by construction. As a result, the solutions would have the desired monodromy structure around $z=\infty$ as well when we think this contour to surround $z=\infty$ instead. So we conclude that the solutions of the Fuchsian equation (\ref{eq:Fuchplugged}), with the right choice of $v(\lambda_1,\lambda_2, \lambda_{3})$, can realize hyperbolic SL(2$, \mathbb{R}$) monodromies around each puncture and they are compatible. We solved the monodromy problem.

Finally, we can list the scaled ratios $\rho_i = (\psi_i^+(z)/\psi_i^-(z))^{1/i \lambda_i}$ associated with each puncture. They are:
\begin{align}
\rho_1(z) &= e^{\frac{v(\lambda_1,\lambda_2, \lambda_3)}{\lambda_1}} z (1-z)^{-\frac{\lambda_2}{\lambda_1}} 
\left[ \frac{_2F_1 \left(\frac{1+i \lambda_1 -i \lambda_2 +i \lambda_3}{2},\frac{1+i \lambda_1 -i \lambda_2 -i \lambda_3}{2}; 1+ i \lambda_1; z\right)}{_2F_1 \left(\frac{1-i \lambda_1 +i \lambda_2 -i \lambda_3}{2},\frac{1-i \lambda_1 +i \lambda_2 +i \lambda_3}{2}; 1- i \lambda_1; z\right)} \right]^{\frac{1}{i \lambda_1}}, \nonumber\\
\rho_2(z) &= e^{\frac{v(\lambda_2,\lambda_1, \lambda_3)}{\lambda_2}} (1-z) z^{-\frac{\lambda_1}{\lambda_2}} ,
\left[ \frac{_2F_1 \left(\frac{1+i \lambda_2 -i \lambda_1 +i \lambda_3}{2},\frac{1+i \lambda_2 -i \lambda_1 -i \lambda_3}{2}; 1+ i \lambda_2; 1-z\right)}{_2F_1 \left(\frac{1-i \lambda_2 +i \lambda_1 -i \lambda_3}{2},\frac{1-i \lambda_2 +i \lambda_1 +i \lambda_3}{2}; 1- i \lambda_2; 1-z\right)} \right]^{\frac{1}{i \lambda_2}},\nonumber \\
\rho_3(z) &= e^{\frac{v(\lambda_3,\lambda_2, \lambda_1)}{\lambda_3}} \left(\frac{1}{z}\right) \left(1-\frac{1}{z}\right)^{-\frac{\lambda_2}{\lambda_3}} 
\left[ \frac{_2F_1 \left(\frac{1+i \lambda_3 -i \lambda_2 +i \lambda_1}{2},\frac{1+i \lambda_3 -i \lambda_2 -i \lambda_1}{2}; 1+ i \lambda_3; \frac{1}{z}\right)}{_2F_1 \left(\frac{1-i \lambda_3 +i \lambda_2 -i \lambda_1}{2},\frac{1-i \lambda_3 +i \lambda_2 +i \lambda_1}{2}; 1- i \lambda_3; \frac{1}{z}\right)} \right]^{\frac{1}{i \lambda_3}}. \label{eq:scaledratio}
\end{align}
From above it is clear that $\rho_2(z)$ and $\rho_3(z)$ can be obtained from $\rho_1(z)$ by exchanging punctures, as well as their associated $\lambda_j$'s, $(1) \leftrightarrow (2)$ and $(1) \leftrightarrow (3)$ respectively while keeping the remaining puncture fixed. Moreover, one can also show that the scaled ratio associated with the fixed puncture remains invariant (up to a sign) under this exchange, either by reasoning through our construction above or by checking it directly using the identities for hypergeometric functions~\cite{hypergeometric}. In any case, we see that the set of three scaled ratios given above would be invariant (up to a sign) under the permutation group $S_3$ acting on the positions and the parameters of the punctures. This fact will eventually lead us to a similar symmetry for the local coordinates of the hyperbolic three-string vertex.

As we already argued in the previous subsection, these scaled ratios will define the following single-valued, singular, hyperbolic metric on the whole three-punctured sphere (\ref{eq:met}):
\begin{equation} \label{eq:yetanothermetric}
ds^2 = \frac{\lambda_{j}^2 |\partial \rho_j (z)|^2}{|\rho_j(z)|^2 \sin^2(\lambda_j \log|\rho_j(z)|)} |dz|^2
=  \frac{\lambda_{j}^2 }{|\rho_j|^2 \sin^2(\lambda_j \log|\rho_j|)} |d\rho_j|^2,
\end{equation}
for which we have three semi-infinite series of attached hyperbolic cylinders connected to each other. Again, each $j=1,2,3$ defines the same metric.

\subsection{The resulting geometry on the three-punctured sphere}

\begin{figure}[!pht]
	\centering
	\fd{14cm}{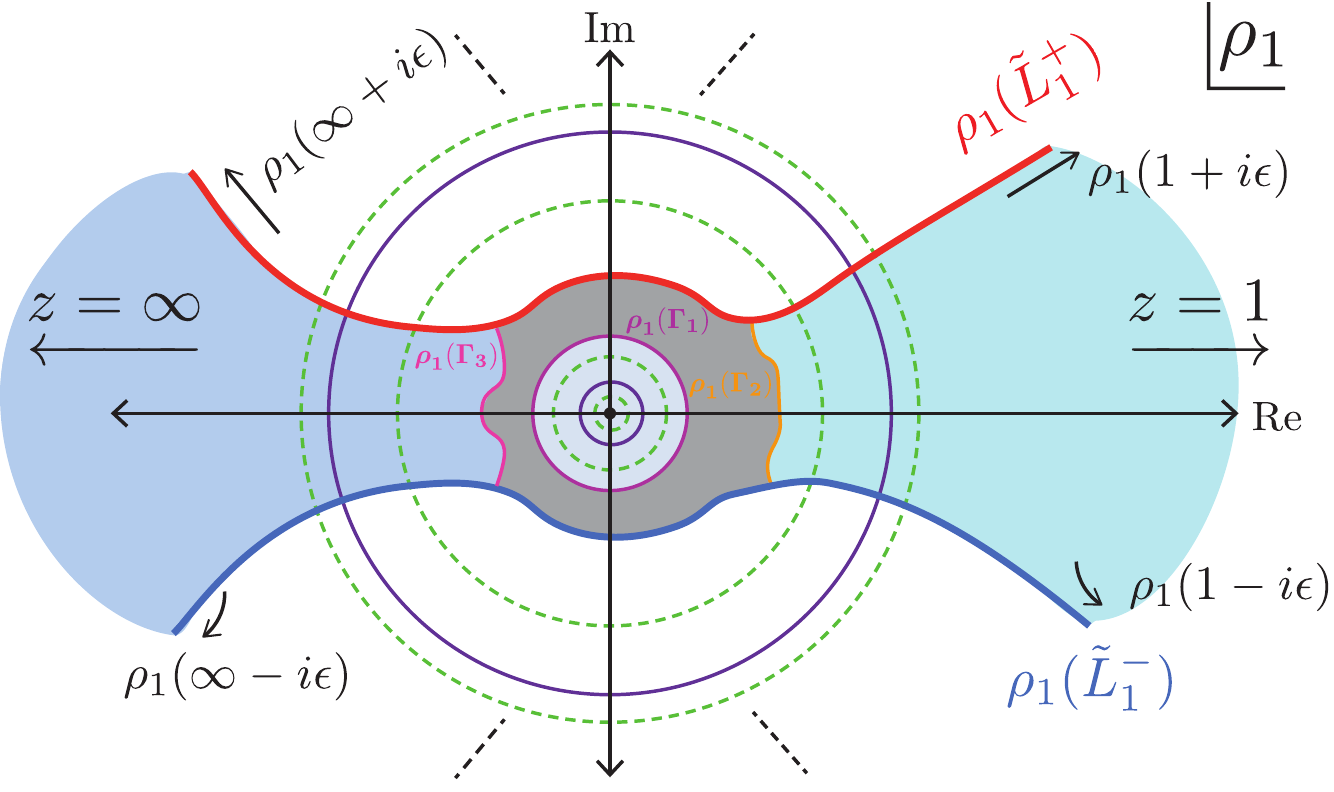}
	\caption{The rough sketch of the geometry on the $\rho_1$-plane. The meaning of the curves are explained in the text. The coloring conventions for the curves will be the same for all figures in this subsection. Dashed curves indicate the line singularities. \vspace{1cm}}
	\label{fig:rhoSketch}
	\centering
	\fd{14cm}{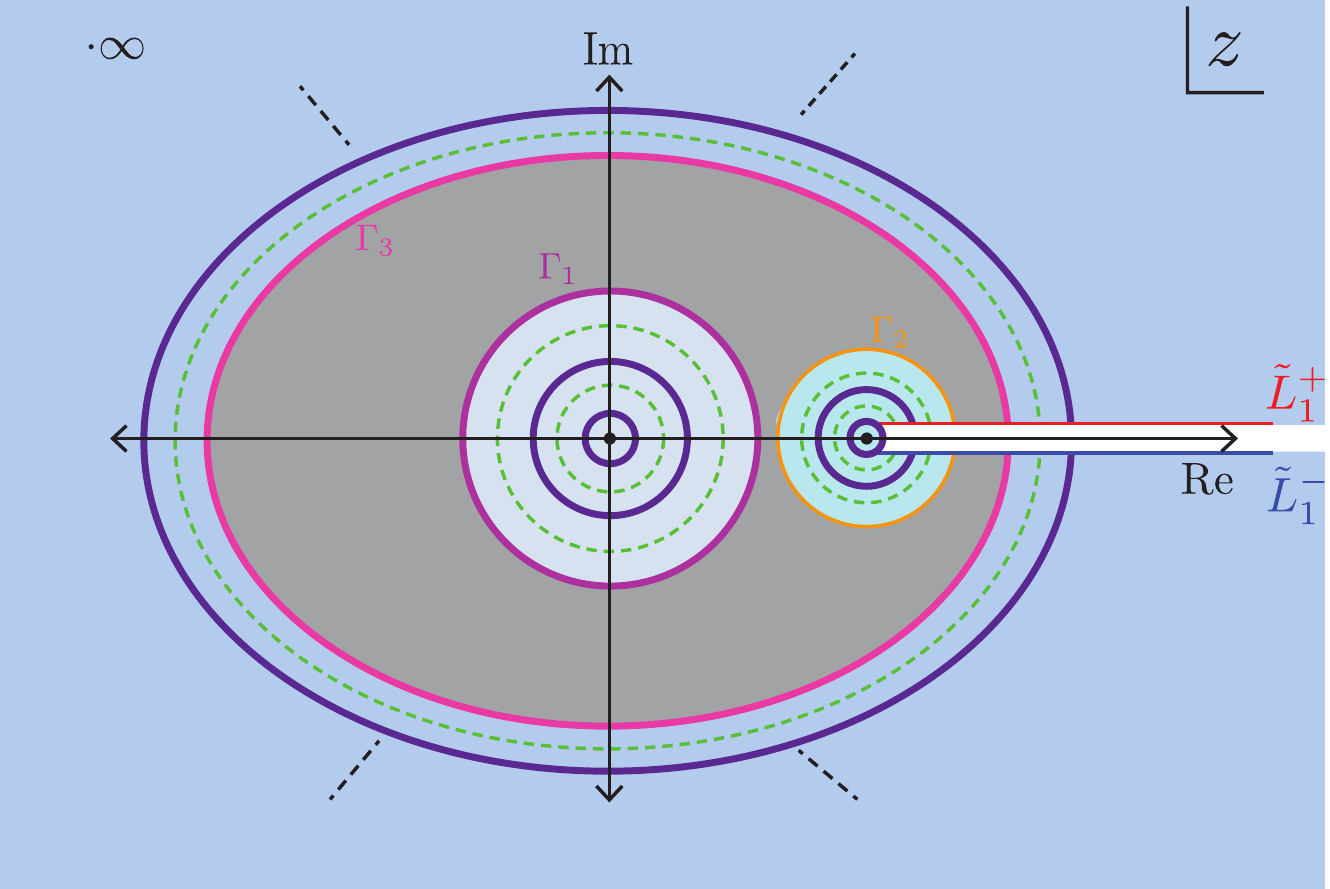}
	\caption{The corresponding geometry on the $z$-plane after we pullback the metric~(\ref{eq:yetanothermetric}) from the $\rho_1$-plane above. Note that the gray region would be endowed with the hyperbolic metric with geodesic boundaries $\Gamma_i$.}
	\label{fig:zSketch}
\end{figure}

Before we construct the local coordinates, we should understand the geometry of (\ref{eq:yetanothermetric}) better and show that it looks exactly like in figure \ref{fig:CylinderSketch} as we have claimed. In order to do that, focus on the set $S_1$, which was the complex plane with a cut from $1$ to $\infty$ (see~\eqref{eq:s1}). This will be mapped to the set $\rho_1(S_1)$ in the $\rho_1$-plane.\footnote{It can be shown that this map is invertible, see \cite{hypergeometric}. So this mapping would be bijective.} The rough sketch of these regions, based on numerics, but \emph{not} on scale, is given in figure \ref{fig:rhoSketch} and \ref{fig:zSketch}. We will consider and explain this geometry on the $\rho_1$-plane for now, but geometries on the other $\rho_j$-planes are analogous.

As we mentioned previously, the metric on the $\rho_1$-plane (\ref{eq:yetanothermetric}) takes the form of the hyperbolic metric of series of attached hyperbolic cylinders. Indeed, we see that the line singularities (where the metric blow up on a curve) and the simple closed geodesics surrounding the origin $\rho_1=0$ of the hyperbolic metric~\eqref{eq:yetanothermetric} are located at
\begin{equation}
\text{Line singularities:} \; |\rho_1| = e^{\frac{\pi l_1}{\lambda_1}}, \qquad
\text{Simple Closed Geodesics:} \;  |\rho_1| = e^{\frac{\pi}{\lambda_1} \left(l_1+\frac{1}{2}\right)},
\end{equation}
where $ l_1 \in \mathbb{Z}$. We can see these by noting that the sine in the denominator of the metric~\eqref{eq:yetanothermetric} is equal to zero in the case of line singularity  by $\sin(\pi l_1) = 0$ and one in the case of simple closed geodesics by $\sin\left(\pi l_1+ \pi/2 \right ) = 1$, which makes the metric~\eqref{eq:yetanothermetric} blow up and minimize respectively.

Notice that the line singularities and simple closed geodesics form alternating, exponentially separated circles around the origin on the $\rho_1$-plane, as shown in figure \ref{fig:rhoSketch} with green and purple respectively; except for the geodesic colored with magenta which will turn out to be special. Additionally, it is clear that every simple geodesic surrounding the origin has the length $2 \pi \lambda_1$ by the metric (\ref{eq:yetanothermetric}), which justifies the name \emph{geodesic radii} for $\lambda_j$. Obviously we can pullback these curves to the $z$-plane with a cut from 0 to $\infty$, which will result in closed, simple geodesics/line singularities around the puncture $z=0$ by $\rho_1(0)=0$. These are shown in figure~\ref{fig:zSketch} correspondingly.

Observe that the lines just above/below the branch cut of $\rho_1(z)$, denoted as $\widetilde{L}_1^{\pm}$ and shown in figure \ref{fig:zSketch}, are mapped to the red/blue curves $\rho_1 (\widetilde{L}_1^{\pm})$ in $\rho_1(S_1)$. These curves are shown in figure \ref{fig:rhoSketch}. They are symmetric with respect to the real axis on the $\rho_1$-plane by the choice of the principal branch for the scaled ratio. The set $S_1$ is mapped between $\rho_1(\widetilde{L}_1^+)$ and  $\rho_1(\widetilde{L}_1^-)$, which is the shaded region in figure \ref{fig:rhoSketch}. Moreover, if we identify the two curves $\rho_1 (\widetilde{L}_1^{\pm})$, the whole $z$-plane minus the punctures maps to the region between them. But, in any case, we indicated where the punctures $z=1$ and $z= \infty$ are heuristically getting mapped to in figure \ref{fig:rhoSketch}: $z=1$ is mapped to the right-side infinity and $z= \infty$ is mapped to the left-side infinity. 

Now observing figure \ref{fig:rhoSketch}, we see that some simple closed geodesics/line singularities don't intersect $\rho_1(\widetilde{L}^{\pm}_1)$. As a result, we see $\exists \, \tilde{l}_1 \in \mathbb{Z}$ such that the geodesic at $|\rho_1| = e^{\frac{\pi}{\lambda_j}\left( \tilde{l}_1+\frac{1}{2}\right)}$ does not intersect $\rho_1(\widetilde{L}^{\pm}_1)$ and surrounds \emph{all} the closed simple geodesics/line singularities that do not intersect $\rho_1(\widetilde{L}^{\pm}_1)$ (i.e. those with $l_1 \leq \tilde{l}_1$). The closed geodesic with $l_1 = \tilde{l}_1$ is shown with magenta instead of purple in figure~\ref{fig:rhoSketch} in order to differentiate it from the others. At this stage nothing prevents us to having a line singularity that surrounds this geodesic and doesn't intersect $\rho_1(\widetilde{L}^{\pm}_1)$, but this turns out not to be the case as we will prove it shortly. We just assume this is the case for now.

We can pullback the geodesic with $l_1=\tilde{l}_1$ described above to the $z$-plane, which we denote it by $\Gamma_1$. Defining the closed geodesics homotopic to the puncture $z=0$ as \emph{separating} geodesics of $z=0$, we see the simple closed geodesic $\Gamma_1$ would be the separating geodesic farthest away from $z=0$ by construction. So we will call $\Gamma_1$ as \emph{the most-distant separating} geodesic of $z=0$. This geodesic is shown in figure \ref{fig:zSketch} with magenta as well.

From this, we see that the simply-connected region $H_1$ on the $z$-plane surrounded by $\Gamma_1$ contains every geodesic/line singularity with $l_1 \leq \tilde{l}_1$. Furthermore, as $l_1 \to - \infty$, the geodesics/line singularities get closer to the puncture. So we conclude that the geometry on $H_1$ looks like a series of semi-infinite hyperbolic cylinders attached at where they flare up, like shown in figure~\ref{fig:CylinderSketch}. The places where they flare up are the line singularities of the metric.

We can repeat the same procedure for the other punctures and obtain their most-distant separating geodesics $\Gamma_j$, associated simply-connected regions $H_j$, and integers $\tilde{l}_j$. Note that $\Gamma_i \cap \Gamma_j = \emptyset$ for $i \neq j$, by $\Gamma_i$'s being simple geodesics of the same metric. Hence, the resulting geometry on the $z$-plane would indeed look like in figure \ref{fig:zSketch}. Again, the most-distant separating geodesics $\Gamma_j$ are shown with different colors. In this figure, we also see there are alternating closed curves around each puncture representing the simple closed geodesics/line singularities surrounding them. These can be related to the geodesics/line singularities that intersect $\rho_1 (\widetilde{L}_1^{\pm})$ on the $\rho_1$-plane (hence their colors), but this wouldn't be necessary for our purposes.

Now let us inspect how the most-distant separating geodesics of the punctures $z=1$ and $z=\infty$, denoted as $\Gamma_2$ and $\Gamma_3$ respectively, look like on $\rho_1(S_1)$. In order to do that, let us call the line singularity with $l_1 = \tilde{l}_1 + 1$ to be the \emph{first line singularity} of $z=0$. Clearly, the first line singularity encloses $\rho_1(\Gamma_1)$ and is enclosed by every other line singularity that encloses $\rho_1(\Gamma_1)$ on the $\rho_1$-plane, hence the name \emph{first}. Moreover, it is clear that the first line singularity intersects with the curves $\rho_1(\widetilde{L}_1^{\pm})$ by definition. We define the \emph{first geodesic} of a puncture in similar fashion.

Now, we will find the shortest geodesic that is enclosed by the first line singularity of $z=0$ and stretches between the curves $\rho_1(\widetilde{L}_1^{\pm})$ for both right/left of the origin $\rho_1 =0$ on $\rho_1(S_1)$, which we will call $\rho_1(\Omega_2)$ and $\rho_1(\Omega_3)$ respectively. Clearly, $\rho_1(\Omega_2)$ and $\rho_1(\Omega_3)$ can be made shorter by eliminating any self intersections, so we will consider the simple geodesics without loss of generality. Moreover, $\rho_1(\Omega_2)$ and $\rho_1(\Omega_3)$ can be made shorter by making them intersect $\rho_1(\widetilde{L}_1^{\pm})$ perpendicularly, which we will also take to be the case.

There might be multiple curves satisfying the definition for $\rho_1(\Omega_2)$ and $\rho_1(\Omega_3)$ above. However, this cannot be the case since their pullbacks on the $z$-plane would correspond to closed simple geodesics without a line singularity between them around the punctures $z=0$ and $z=\infty$, and we know that this can't happen as we saw above. So $\rho_1(\Omega_2)$ and $\rho_1(\Omega_3)$ are unique for the left and right side. This is shown in figure~\ref{fig:rhoSketch}. Additionally, this argument shows that $\Omega_2$ and $\Omega_3$ are the the most-distant separating geodesics for the punctures $z=1$ and $z=\infty$ respectively, i.e. $\Gamma_2 = \Omega_2$ and $\Gamma_3=\Omega_3$, since there are no geodesics that surround them and separate from the other punctures.

\begin{figure}[!t]
	\centering
	\fd{7.5cm}{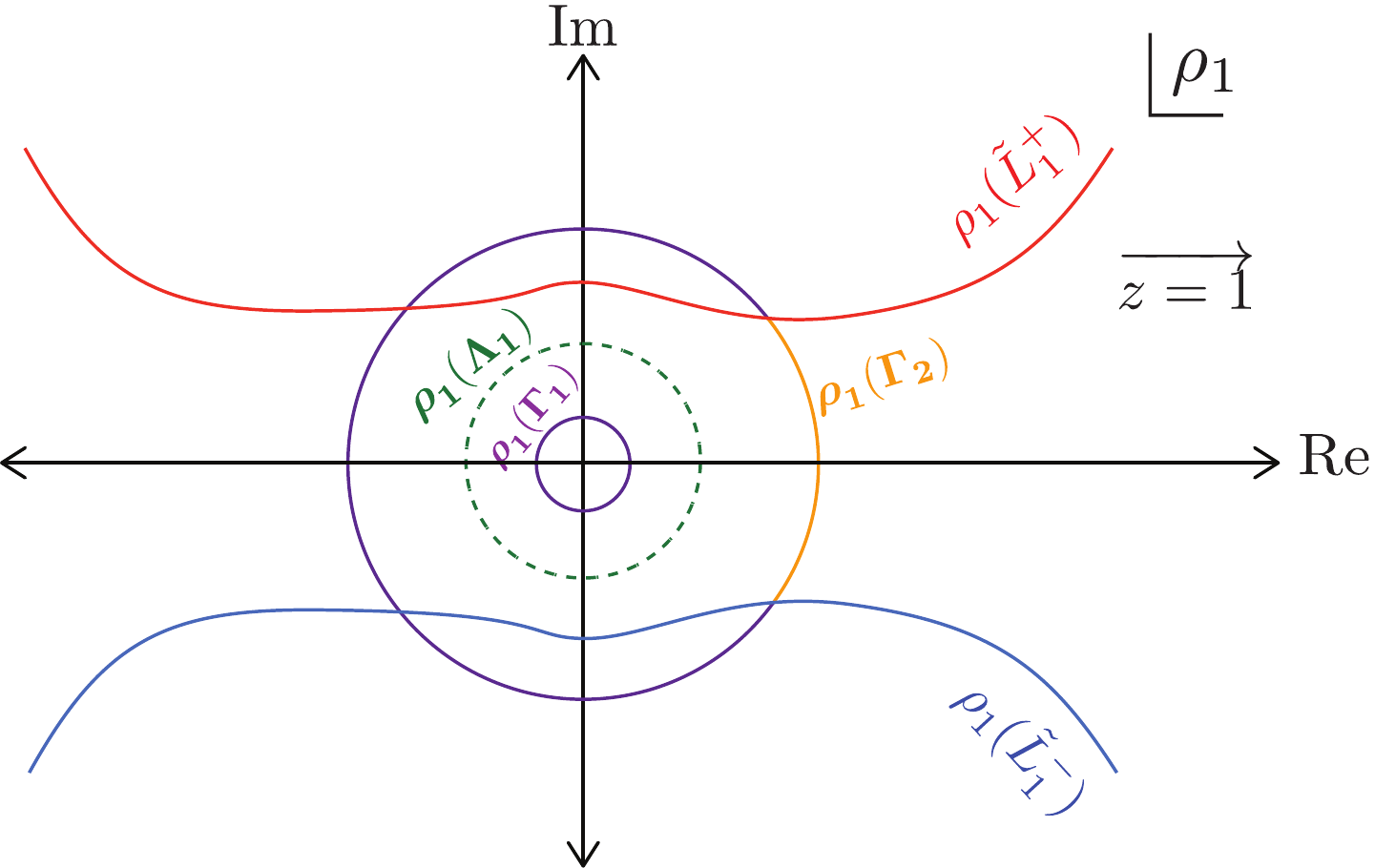}
	\fd{7.5cm}{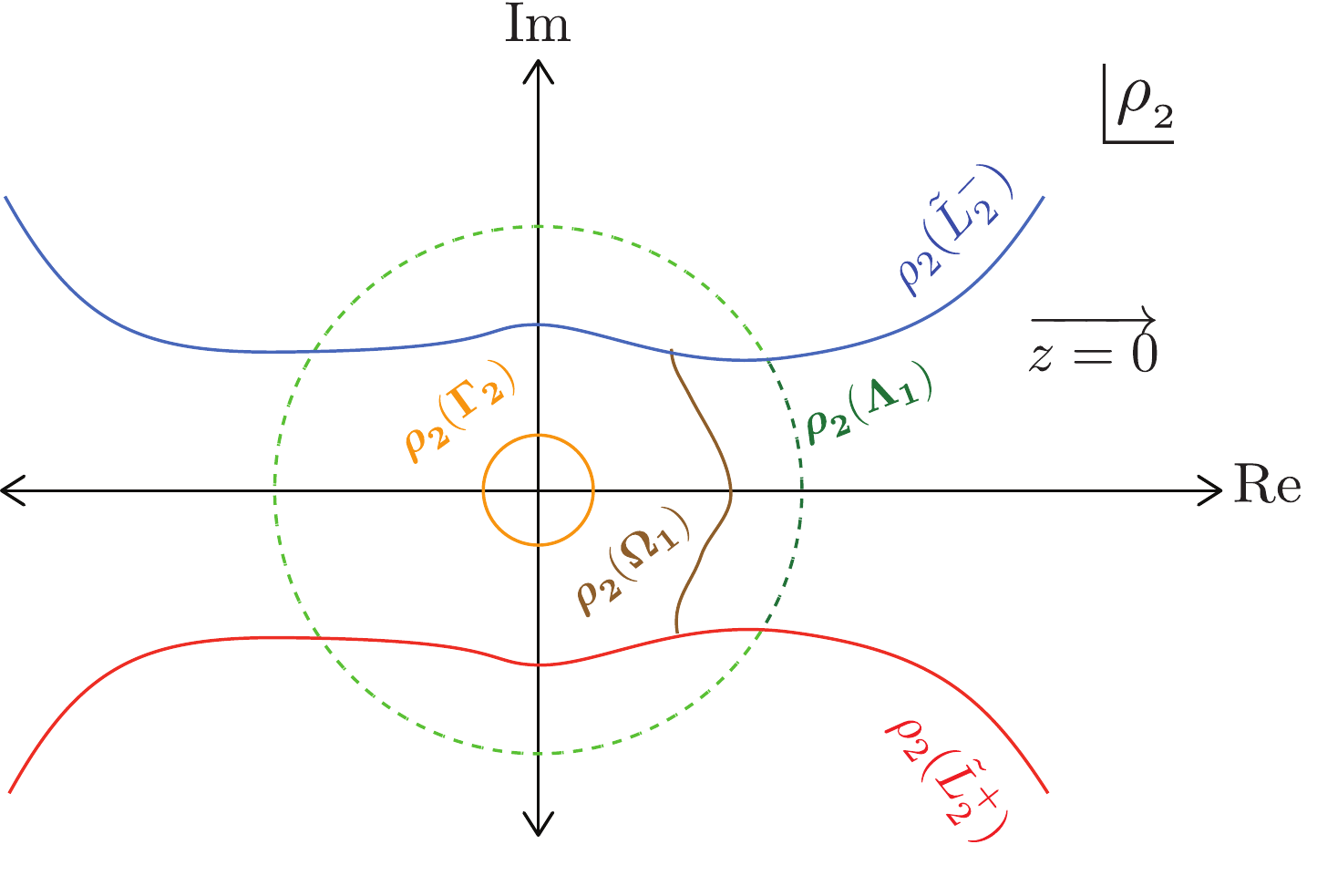}
	\caption{The illustration of the geometry described in the argument above on the $\rho_1$-plane (left) and $\rho_2$-plane (right). Here the line singularity $\Lambda_1$ is shown with dark green, while the geodesic $\Omega_1$ is shown with brown.}
	\label{fig:argument}
	\centering
\end{figure}

Keeping this in mind, we can now demonstrate that the there is no line singularity that surrounds the geodesic $\rho_1(\Gamma_1)$ and doesn't intersect $\rho_1(\widetilde{L}^{\pm}_1)$ on the $\rho_1$-plane, which we only assumed previously. For the sake of contradiction, suppose there is one and call it $\rho_1(\Lambda_1)$, which is shown in figure~\ref{fig:argument}. Then it is clear by above that the geodesic $\rho_1(\Gamma_2)$ around the puncture $z=1$ would be a piece of the first geodesic of $z=0$. Now going to the $\rho_2$-plane after we pullback this geometry to the $z$-plane, we see that $\Lambda_1$ maps to a piece of a line singularity $\rho_2(\Lambda_1)$ on the $\rho_2$-plane stretching between $\rho_2(\widetilde{L}_2^{\pm})$, while $\Gamma_2$ maps to a circle around the origin and doesn't intersect $\rho_2(\widetilde{L}_2^{\pm})$. Similar to the arguments above, we can always find a simple geodesic $\rho_2(\Omega_1)$ between these two, but this leads to a contradiction with the fact that $\Gamma_1$ being the most-distant separating geodesic of $z=0$ since the separating geodesic $\Omega_1$ would be enclosing $\Gamma_1$. Clearly this argument can be repeated for other punctures, so what we have assumed regarding having a line singularity that surrounds the most-distant separating geodesic and doesn't intersect $\rho_1(\widetilde{L}^{\pm}_1)$ was justified.

In order to complete our construction, we now need to find the integers $\tilde{l}_j$.  For that, first notice the following inequality is satisfied:
\begin{equation} \label{eq:ineq}
|\rho_j(z)| = \exp \left[\frac{\pi}{\lambda_j}\left( \tilde{l}_j+\frac{1}{2}\right)\right] \leq \min_{z \in \widetilde{L}_j} |\rho_j(z)|,
\end{equation}
with $\widetilde{L}_j$ denoting the branch cut of the function $\rho_j(z)$. This inequality is evident since we demanded above that the geodesic $\rho_j(\Gamma_j)$, located at $|\rho_j| = e^{\frac{\pi}{\lambda_j}\left( \tilde{l}_j+\frac{1}{2}\right)}$, is not intersecting the curves $\rho_j(\widetilde{L}^{\pm}_1)$ on the $\rho_j$-plane. Note that $|\rho_j(z)|$ would be single-valued on the branch cut $\widetilde{L}_j$ because of the choice of the principal branch. From~\eqref{eq:ineq} and noting that $\tilde{l}_j$ is the greatest integer that satisfies it by definition, we can write a prescription for $\tilde{l}_j$ as follows:
\begin{align} \label{eq:ltilde}
\tilde{l}_j = \left\lfloor \frac{\lambda_j}{\pi}  \log \min_{z \in \widetilde{L}_j} |\rho_j(z)| - \frac{1}{2}\right\rfloor.
\end{align}
Here $\lfloor \cdot \rfloor: \mathbb{R} \to \mathbb{Z}$ denotes the floor function. We couldn't be able to find an explicit expression for this in terms of $\lambda_j$'s. However, determining the \emph{exact} values of the integers $\tilde{l}_j$ numerically for given $\lambda_j$ is trivial by the expression above and using the scaled ratios~\eqref{eq:scaledratio}. 

Although it is hard to find an expression for $\tilde{l}_j$ in terms of arbitrary $\lambda_j$'s, we can still make some progress for the case where two of the $\lambda_j$'s are equal by exploiting the permutation symmetry. In order to do that, suppose we want to find $\tilde{l}_1$ in the case of $\lambda_2=\lambda_3 = \lambda$. Now recall that three scaled ratios~\eqref{eq:scaledratio} are invariant under the permutations of the punctures and their associated geodesic radii up to a sign. Specifically, in the case where we exchange the punctures at $z=1,\infty$ while keeping $z=0$ fixed, which is implemented by the conformal transformation $z \to \frac{z}{z-1}$, we get the following relation for $\rho_1(z)$ on $S_1$
\begin{equation}
\rho_1(z) = -\rho_1 \left(\frac{z}{z-1}\right).
\end{equation} 
Note that it was essential to take $\lambda_2 = \lambda_3$ to establish this relation.

Clearly, $z=2$ is the fixed point of the transformation $z \to \frac{z}{z-1}$. One consequence of this is that $|\rho_1(z)|$ when restricted to the branch cut $\widetilde{L}_1$ is symmetric around $z=2$ when we apply the transformation $z \to \frac{z}{z-1}$. Then using this fact and analyticity of $\rho_1(z)$, it can be shown that the point $z=2$ would be where $|\rho_1(z)|$ attains its global minimum on the branch cut.

\begin{figure}[!t]
	\centering
	\fd{12cm}{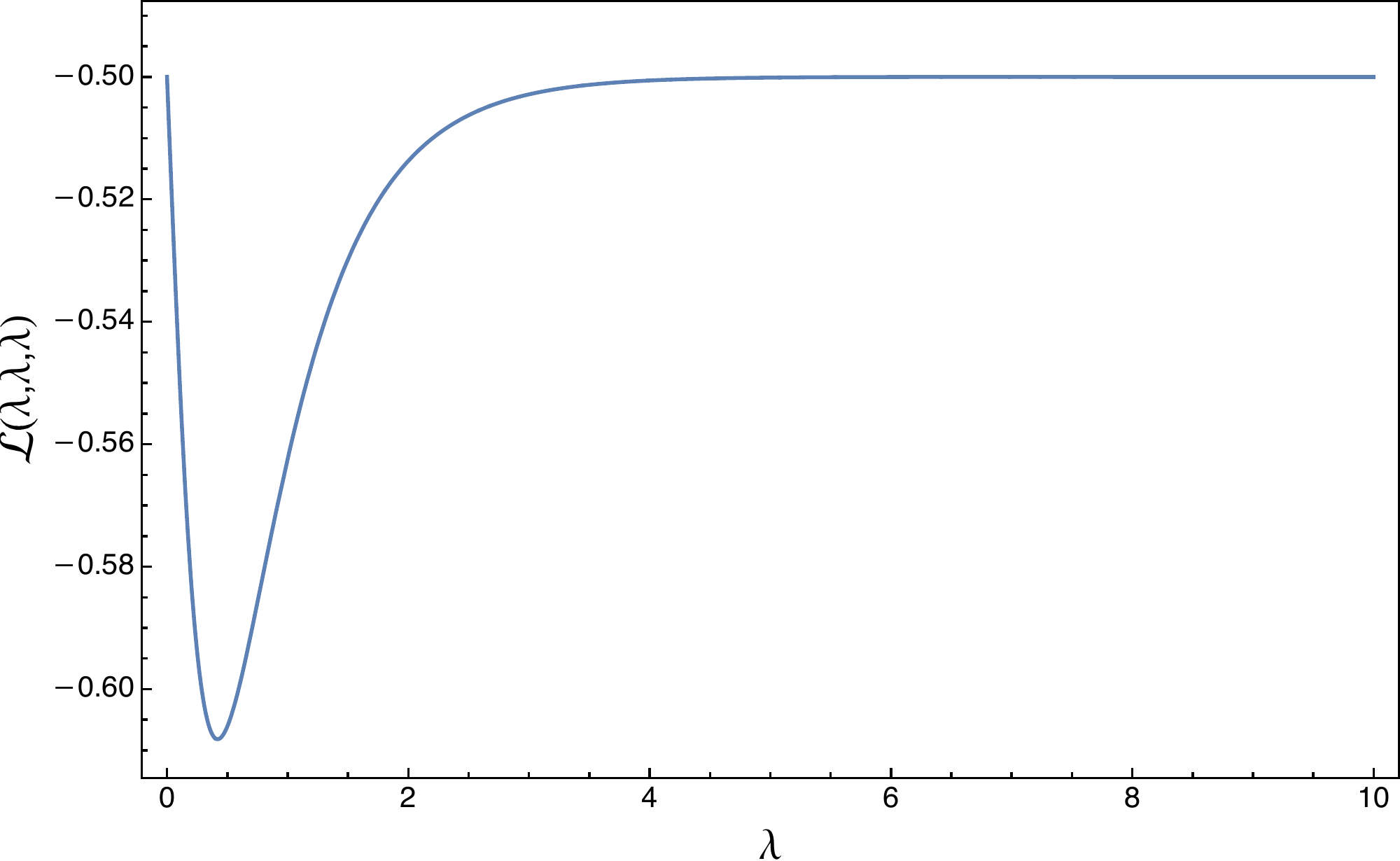}
	\caption{The plot of $\mathcal{L}(\lambda,\lambda,\lambda)$ as a function of $\lambda$. Note that the floor of this function gives $-1$ in the range shown. An analytic proof for $\tilde{l}_1 = -1$ for any $\lambda>0$ would require better understanding of $\mathcal{L}(\lambda,\lambda, \lambda) $, especially for the large values of $\lambda$.}
	\label{fig:Length}
\end{figure}
So we see that permutation symmetry of the situation $\lambda_2=\lambda_3 = \lambda$ allow us to find the global minimum of $|\rho_1(z)|$ on its branch cut $\widetilde{L}_1$, which is at $z=2$.  Now define the following function and notice
\begin{equation}
\mathcal{L}(\lambda_1,\lambda, \lambda) \equiv \frac{\lambda_1}{\pi}  \log |\rho_1(2)| - \frac{1}{2} \implies \tilde{l}_1 = \lfloor \mathcal{L}(\lambda_1,\lambda, \lambda) \rfloor,
\end{equation}
using (\ref{eq:ltilde}). This expression is certainly more manageable then what has been given in~(\ref{eq:ltilde}). Obviously, we can get similar expressions for the other punctures when the remaining punctures has equal $\lambda_j$'s. As an example for what we have discussed so far, we plotted $\mathcal{L}(\lambda,\lambda, \lambda)$ in figure~\ref{fig:Length}. This suggests $\tilde{l}_1=-1$, and by symmetry $\tilde{l}_2 = \tilde{l}_3=-1$, for  $0 < \lambda < 10$.

\begin{figure}[!t]
	\centering
	\fd{7.5cm}{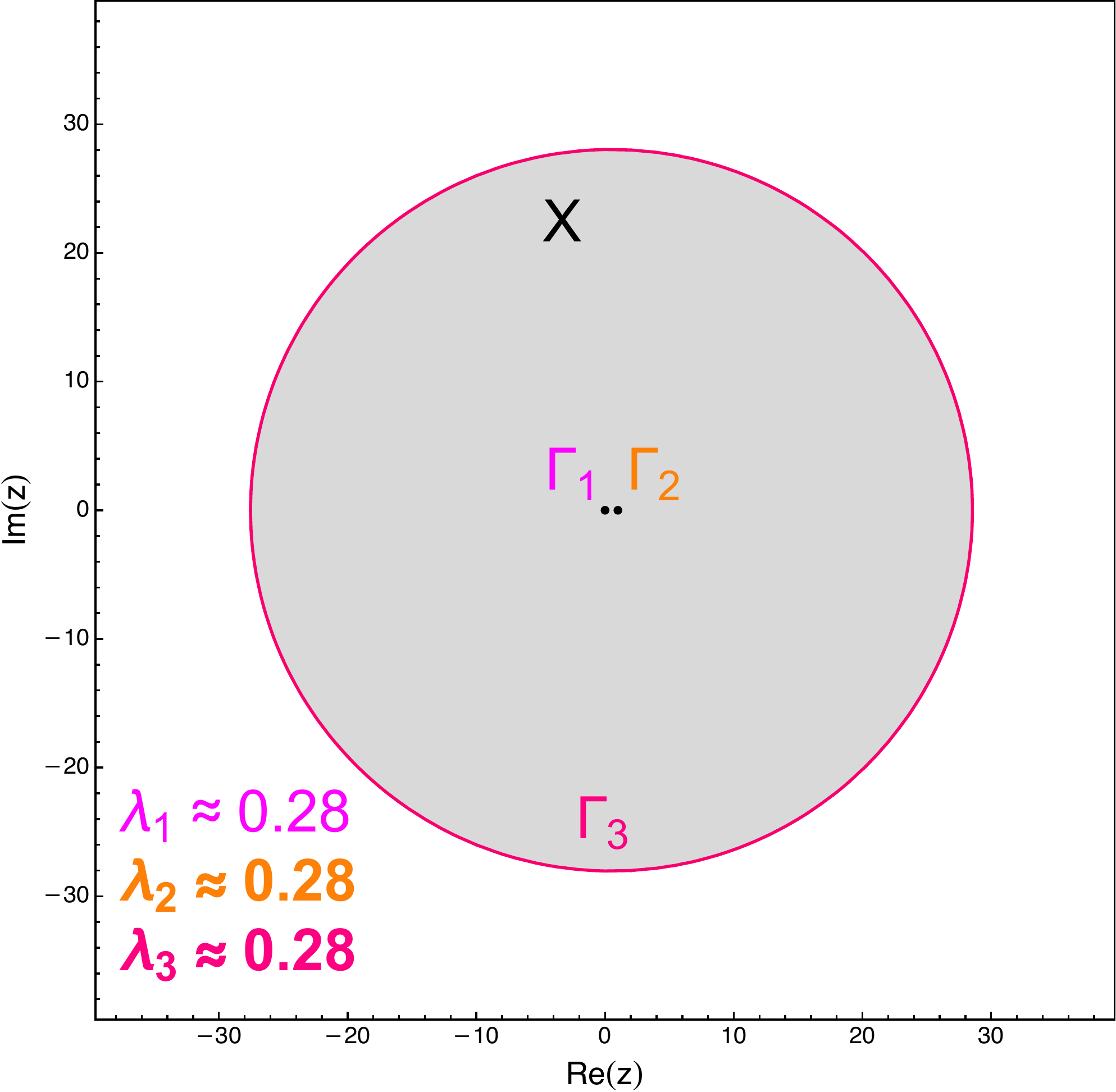}
	\fd{7.5cm}{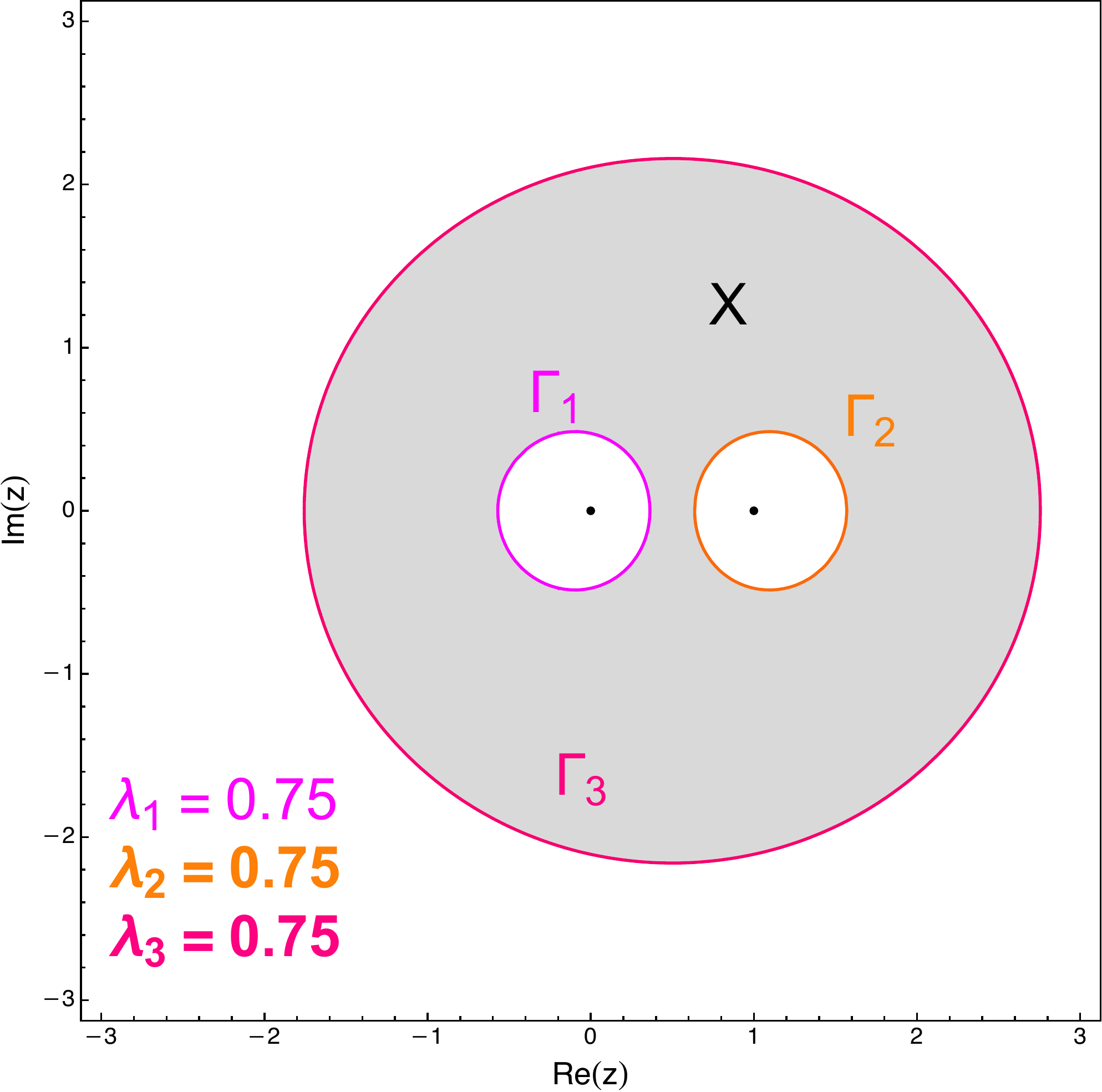}
	\fd{7.5cm}{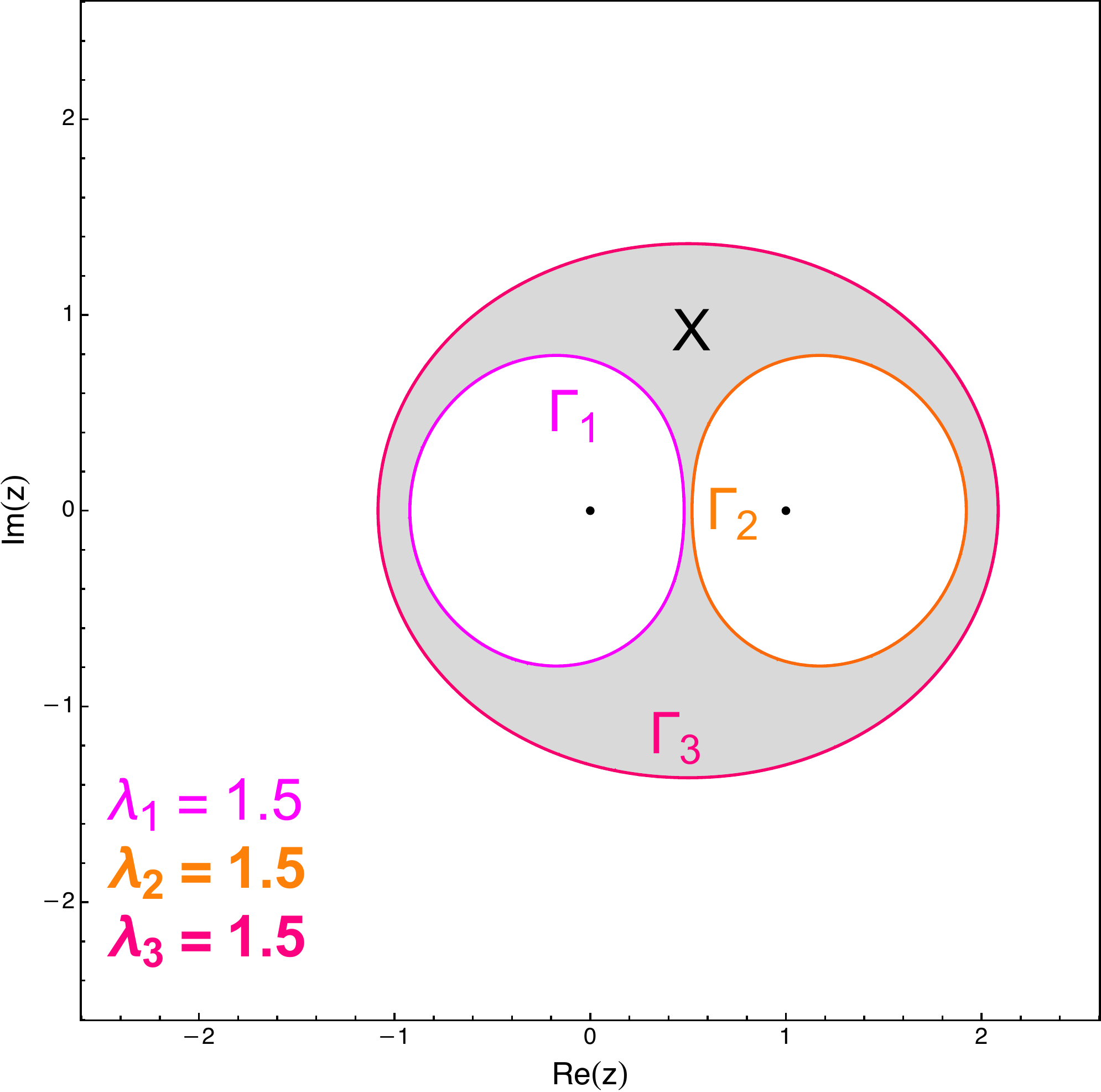}
	\fd{7.5cm}{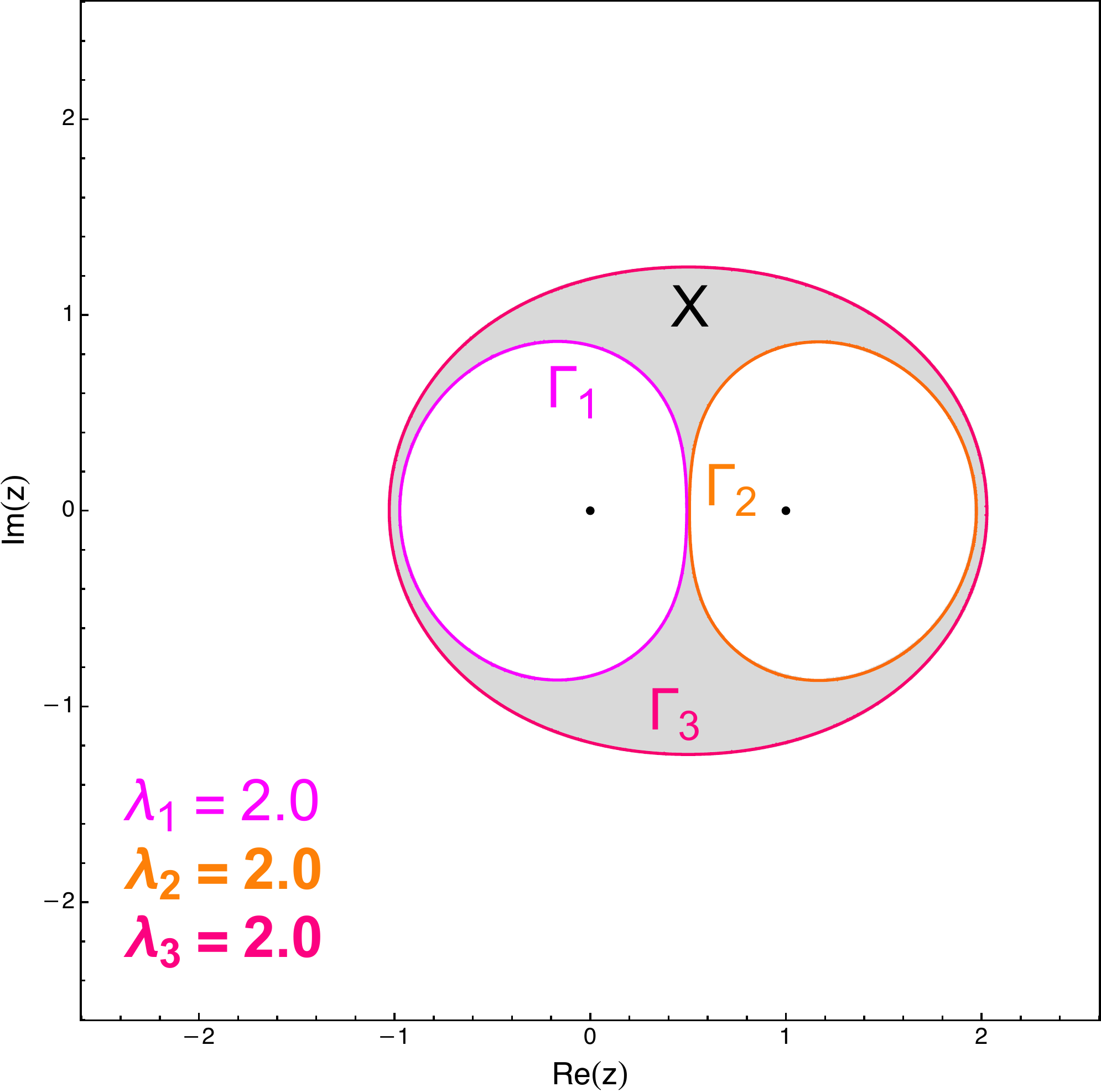}
	\caption{Four examples for the pants diagram region $X$ on the $z$-plane in the case of equal geodesic radii. Punctures are located at $z=0,1,\infty$, and indicated by black dots. We only show the most-distant separating geodesics $\Gamma_j$ because we performed a surgery and take out everything surrounded by them. The region remaining $X$ is endowed with the hyperbolic metric (\ref{eq:yetanothermetric}) and $\Gamma_j$ are its geodesics by construction. Note that the geodesics $\Gamma_1$ and $\Gamma_2$ in the critical case $\lambda =$ arcsinh(1)$/\pi \approx 0.28$ are so small that they haven't rendered in the top-left figure.}
	\label{fig:example}
\end{figure}

In summary, we see the geometry of the metric~(\ref{eq:yetanothermetric}) is indeed given by figure \ref{fig:CylinderSketch}. Remember the metric (\ref{eq:yetanothermetric}) was on the three-punctured sphere $\widetilde{X}$, but, clearly, we can now obtain the hyperbolic metric with geodesic boundaries on a three-holed sphere by restricting to the region $X=\widehat{\mathbb{C}} \setminus (H_1 \cup H_2 \cup H_3)$, which is shaded gray in figure \ref{fig:zSketch}. On the $\rho_j$-plane this corresponds to the region between the most-distant geodesics with the curves $\rho_j(L^{\pm}_j)$ are identified, which is also shaded gray in figure \ref{fig:rhoSketch}. Note that $X$ is still endowed with $K=-1$ metric (\ref{eq:yetanothermetric}), but now free from singularities, and it is clear by the construction that its boundaries $\Gamma_j$ are geodesics. In other words, we performed a \emph{surgery} where we amputated the hyperbolic cylinders around the hyperbolic singularities and left with the geodesic boundaries instead while keeping everything the same. Four examples of such region on the $z$-plane are shown in figure~\ref{fig:example}. In the next subsection, we are going to graft flat semi-infinite cylinders into the places of amputated hyperbolic cylinders in order to construct the local coordinates explicitly.

\subsection{Local coordinates} \label{sec:Local}
In this subsection, we describe how to construct the local coordinates around the punctures for the hyperbolic three-string vertex by attaching flat semi-infinite cylinders of radius $\lambda_j$ at each geodesic boundary component of $X$. First, note that when we perform the surgery described above to obtain the geodesic boundaries, we essentially take out the disk 
\begin{equation}
\rho_j(H_j) = \left\{\rho_j \in \mathbb{C} \; | \; |\rho_j| < \exp \left[ \frac{\pi}{\lambda_j} \left(\tilde{l}_j+\frac{1}{2} \right) \right]\right\} ,
\end{equation}
from the $\rho_j$-plane for each $j=1,2,3$. Now imagine we have a punctured unit disk $0 < |w_j| \leq 1$ with the metric
\begin{equation} \label{eq:cylmetric}
ds^2 = \lambda_j^2 \frac{ |dw_j|^2}{|w_j|^2},
\end{equation}
which describes a flat semi-infinite cylinder ($K=0$) of radius $\lambda_j$. We can map this punctured unit disk into the hole $\rho_j(H_j)$ on the $\rho_j$-plane with a simple scaling:
\begin{equation} \label{eq:scale}
\rho_j =  \exp \left[ \frac{\pi}{\lambda_j} \left(\tilde{l}_j+\frac{1}{2} \right) \right] w_j = N_j w_j, \quad N_j \equiv \exp \left[ \frac{\pi}{\lambda_j} \left(\tilde{l}_j+\frac{1}{2} \right) \right].
\end{equation}
We will call $N_j$ the \emph{scale factor}. Above we haven't considered the overall rotations of the punctured unit disk, $w_j \to e^{i \theta} w_j$, while we are mapping to $\rho_j(H_j)$, since such global phase factors are not relevant in closed string field theory.

Clearly, the flat metric (\ref{eq:cylmetric}) does not change under this scaling. Furthermore, the flat metric~(\ref{eq:cylmetric}) and the hyperbolic metric (\ref{eq:yetanothermetric}) for the pair of pants as well as their first derivatives match at the circular seams $\rho_j(\Gamma_j)$ of radius $\lambda_j$. As a result, we fill the regions  $\rho_j(H_j)$ with flat semi-infinite cylinders and discontinuity first appears in the curvature as we desire. Note that the metric we obtain after grafting these semi-infinite flat cylinders is a Thurston metric on the three-punctured sphere~\cite{Costello:2019fuh}.

Now we can pullback these filled $\rho_j(H_j)$ to the otherwise empty holes $H_j$ on the $z$-plane with the maps $\rho_j(z)$ to construct the local coordinates around the punctures $z=0,1,\infty$ describing three semi-infinite flat cylinders grafted on to the hyperbolic pair of pants on $\widehat{\mathbb{C}}$. Thus, from~\eqref{eq:scale}, we see that the local coordinates $w_j$ around the punctures $w_j=0$ are given by
\begin{equation} \label{eq:LocalCoord}
w_j =   \exp \left[- \frac{\pi}{\lambda_j} \left(\tilde{l}_j+\frac{1}{2} \right) \right] \rho_j(z) = N_j^{-1} \rho_j(z),
\end{equation}
with $|w_j| \leq 1$. This yields the local coordinates~\eqref{eq:lc} using~\eqref{eq:scaledratio}. Equivalently, we can write $z = \rho_j^{-1}\left( N_j w_j\right)$ on the coordinate patches $H_j$ with the punctures are located at $z=z_j$. Note that $|w_j|=1$ maps to $\partial H_j = \Gamma_j$ by construction.  Obviously, we can get the anti-holomorphic local coordinates $w_j(\bar{z})$ in similar fashion. Moreover, we see that they satisfy $w_j(\bar{z}) = \overline{w_j(z)}$ from~\eqref{eq:lc}, up to possible overall phase ambiguity. This shows all the coefficients in the expansions of $w_j(z)$ in $z$ can be chosen to be real.

As can be seen from~\eqref{eq:lc}, and alluded before, the local coordinates are invariant under permutations of the punctures and their associated $\lambda_j$. Adding the scale factor $N_j$ doesn't spoil this symmetry, since its value is getting permuted as well. Moreover, when we take all $\lambda_j = \lambda$ equal (recall this is the version that appears in the string action), this vertex becomes cyclic in the technical sense~\cite{sonoda1990covariant}. These results are certainly consistent with what is expected form the geometry of the hyperbolic pair of pants with three grafted flat cylinders.

As a final note, the mapping radius $r_j = \left| \frac{dz}{dw_j} \right|_{w_j=0}$ for this local coordinates can be easily read from~\eqref{eq:lc}, and they are
\begin{equation}
w_j = e^{-\frac{\pi (\tilde{l}_j+\frac{1}{2})}{\lambda_j}} e^{\frac{v_j}{\lambda_j}} (z-z_j) + \dots \implies r_j = \exp\left[\frac{\pi (\tilde{l}_j+\frac{1}{2})}{\lambda_j} - \frac{v_j}{\lambda_j}\right] = N_j \exp\left[- \frac{v_j}{\lambda_j}\right].
\end{equation}
Remember both $v_j$ and $N_j$ depends on the circumferences of the grafted cylinders as can be seen from~\eqref{eq:v} and~\eqref{eq:scale}.

\section{Limits of the hyperbolic three-string vertex} \label{sec:Limits}

In this section, we investigate various limits of the local coordinates (\ref{eq:LocalCoord}) to check that they are consistent with the literature~\cite{Moosavian:2017qsp,sonoda1990covariant, Zwiebach:1988qp}. We show that it is possible to produce the minimal area three-string vertex, Kleinian vertex, and the light-cone vertex as different limits of the hyperbolic three-string vertex.

\subsection{Minimal area three-string vertex} \label{sec:Witten}

In order to produce the minimal area three-string vertex from the hyperbolic three-sting vertex, we set the lengths of the boundary components of $X$ the same, $\lambda_1=\lambda_{2}=\lambda_{3} =\lambda$, and take $\lambda \to \infty$. Since the lengths of the boundaries of the pair of pants get larger at the same rate while the area of the pair of pants remains constant by the Gauss-Bonnet Theorem in this limit, the pair of pants shrinks and it becomes like a ribbon graph of vanishing width. As a result, after grafting the flat cylinders and rescaling their circumferences, we get the three-vertex obtained from the minimal area metric~\cite{Costello:2019fuh}. Therefore, we see that this is indeed the correct limit to generate the minimal area three-string vertex and we will call it \emph{minimal area limit}. Note that this limiting behavior is also evident from the examples given in figure \ref{fig:example}. We seem to get the usual representation of the minimal area three-string vertex as $\lambda$ gets larger~\cite{Erler:2019loq}. 

In order to consider the minimal area limit explicitly, first notice that we have
\begin{equation}
\lim_{\lambda \to \infty} \exp \left[\frac{v(\lambda,\lambda,\lambda)}{\lambda} \right] = \frac{3\sqrt{3}}{4}.
\end{equation}
This can be obtained from the expression~(\ref{eq:v}) for the function $v(\lambda,\lambda,\lambda)$ and evaluating its limit in Mathematica.

Next, we need to find the limiting value of $N=N_1=N_2=N_3$ in the minimal area limit. Already from figure~\ref{fig:Length} and~\eqref{eq:scale} it can be visually argued that $N \to 1$ as $\lambda \to \infty$, but here we are going to provide an additional heuristic argument why this expectation is correct in case figure~\ref{fig:Length} is misleading in large $\lambda$. To that end, we should first understand the minimal area limit of the hyperbolic metric (\ref{eq:yetanothermetric}). This metric certainly diverges in the minimal area limit, but since we are going to rescale our cylinders at the end, this overall divergence is not a problem. Ignoring this divergence, indicated by $\sim$, the hyperbolic metric, now formally on the ribbon graph of vanishing width, takes the following form in the minimal area limit:
\begin{equation}
ds^2 \sim  \frac{ |\partial \rho_i (z)|^2}{|\rho_i(z)|^2 \sin^2(\infty \log|\rho_i(z)|)} |dz|^2.
\end{equation}
Above $\infty$ in the denominator indicates infinite oscillations of the metric as $\lambda \to \infty$ except for when $|\rho_i(z)|=1$. But note that if we have such infinite oscillations, the metric would certainly be ill-defined. The only time it is well-defined is when we have $|\rho_i(z)|=1$, which produces just a divergence and that is acceptable as we mentioned. Thus, we conclude that the shape of the ribbon graph of vanishing width is described by $|\rho_i(z)|=1$ in the minimal area limit, because this is the only time we have a meaningful limit of the geometry.  Now note that this ribbon graph at $|\rho_i(z)|=1$ can be thought as the union of $\Gamma_i$, which is described by $|\rho_i(z)|=N$, in the minimal area limit by shrinking hyperbolic pair of pants. This gives
\begin{equation}
\lim_{\lambda \to \infty} N = \lim_{\lambda \to \infty} \exp\left[\frac{\pi}{\lambda} \left(\tilde{l}+\frac{1}{2}\right) \right]= 1.
\end{equation}
So our expectation above was indeed correct.

Using the two limits we argued above, we see that the local coordinate around $z=0$~\eqref{eq:lc1} has the following expansion in the minimal area limit:\footnote{We also checked the similar results hold for other punctures. We omit reporting them to avoid repetition.}
\begin{align} \label{eq:WittenExpansion}
w_1 = \frac{3 \sqrt{3}}{4}z &+ \frac{3 \sqrt{3}}{8}z^2 + \frac{27 \sqrt{3} }{64}z^3 + \frac{57 \sqrt{3} }{128}z^4 + \frac{231 \sqrt{3} }{512}z^5 + \frac{459 \sqrt{3} }{1024} z^6+ \frac{7275 \sqrt{3} }{16384}z^7 \nonumber \\ 
&+ \frac{14493 \sqrt{3} }{32768} z^8 + \frac{58077 \sqrt{3} }{131072}z^9 + \frac{116565 \sqrt{3} }{262144}z^{10} + \mathcal{O}(z^{11}).
\end{align}
We obtained this expression by expanding (\ref{eq:lc1}) in $z$ first, then taking the minimal area limit. One can easily observe that the local coordinates around $z=0$ of the minimal area three-string vertex, as given in equation (2.19) of~\cite{sonoda1990covariant} with $a=1$,
\begin{equation} \label{eq:minimalvertex}
z_1 = i\frac{\left(1-\frac{i \sqrt{3} z}{z-2}\right)^{3/2}-\left(1+\frac{i \sqrt{3} z}{z-2}\right)^{3/2}}{\left(1-\frac{i \sqrt{3}
		z}{z-2}\right)^{3/2}+\left(1+\frac{i \sqrt{3} z}{z-2}\right)^{3/2}},
\end{equation}
also has the same expansion (\ref{eq:WittenExpansion}) after an unimportant phase rotation $z_1 \to -z_1$. So, unsurprisingly, these local coordinates match in the minimal area limit.

Comparison was perturbative in $z$ above, however, we think this limiting behavior holds for all orders in $z$. That is $w_1=-z_1$ exactly in the minimal area limit. The best way to show this would be by finding an appropriate asymptotic formula for the hypergeometric function when $\lambda$ is large to generate the expression (\ref{eq:minimalvertex}), similar to the cases given in~\cite{Watson}. In any case, this perturbative analysis would be sufficient for our purposes. In conclusion, we see that the hyperbolic three-string vertex reduces to the minimal are three-string vertex in the limit $\lambda \to \infty$.

\subsection{Kleinian vertex}

Now we consider the opposite limit for which $\lambda_j=\lambda \to 0$. Clearly, the grafted flat cylinders disappears in this limit\footnote{Since this is the case, this naive limit of the local coordinates~\eqref{eq:LocalCoord} seems actually ill-defined. We are going to comment on this point below.} and instead we are left with a purely hyperbolic metric on the three-punctured sphere. So the local coordinates for the hyperbolic three-string vertex is expected to be related to the Kleinian vertex of~\cite{sonoda1990covariant} in this limit, whose local coordinates $z_i$ are given by
\begin{equation} \label{eq:KleinianVertex}
z_1 = e^{i \pi \tau(z)}, \qquad z_2 = e^{-i \pi/\tau(z)}, \qquad z_3 = e^{-i \pi/(\tau(z)\pm 1)}, 
\end{equation}
around the punctures $z=0,1,\infty$ respectively, since it involves the same hyperbolic geometry in its construction which emphasized more recently in~\cite{Moosavian:2017qsp,Moosavian:2017sev}. Here the function $\tau(z)$ is the inverse of the modular $\lambda$-function, which is equal to~\cite{hypergeometric}
\begin{equation} \label{eq:tau}
\tau(z) = i \frac{{_2}F_1(\frac{1}{2},\frac{1}{2},1,1-z)}{{_2}F_1(\frac{1}{2},\frac{1}{2},1,z)} = -\frac{i}{\pi} \log\left( \frac{z}{16} \right) + \mathcal{O}(z).
\end{equation}
We will denote the limit $\lambda_j=\lambda \to 0$ as the \emph{Kleinian limit}.

In order to argue for this limit, first notice that the function $\tau(z)$ satisfies the following equality~\cite{hempel1988uniformization}
\begin{equation}
\{\tau,z\} = \frac{1}{2z^2} + \frac{1}{2(z-1)^2} - \frac{1}{2z(z-1)},
\end{equation}
But recall from~\eqref{eq:Tsect3} and~\eqref{eq:3T} we also have
\begin{equation}
\lim_{\lambda \to 0} \{\rho_j^{i \lambda},z\} = \lim_{\lambda \to 0} T_{\varphi}(z) = \frac{1}{2z^2} + \frac{1}{2(z-1)^2} - \frac{1}{2z(z-1)} ,
\end{equation}
So from these two we immediately conclude
\begin{equation} \label{eq:logroh}
\{\tau,z\} = \lim_{\lambda \to 0} \{\rho_j^{i \lambda},z\} = \{\lim_{\lambda \to 0} \log(\rho_j) ,z\} \implies \lim_{\lambda \to 0} \log(\rho_j(z)) = \frac{a \tau(z) + b}{c \tau(z) + d}.
\end{equation}
Above we moved the limit inside the Schwarzian derivative and used the fact that two equal Schwarzian derivatives must be related to each other by a PGL(2,$\mathbb{C}$) transformation. So here $a,b,c,d \in \mathbb{C}$ and $ad-bc \neq 0$. Note that we can easily determine these constants by expanding both sides of~\eqref{eq:logroh} to leading order in $z$.

Take $z=0$  for instance. We already know the expansion of $\log(\rho_1(z))$ around $z=0$ from~(\ref{eq:ScaledRatio}). In the Kleinian limit this would then yield
\begin{equation}
\lim_{\lambda \to 0} \log(\rho_1(z)) =  \log\left( \frac{z}{16} \right)  + \mathcal{O}(z),
\end{equation}
using properties of Gamma functions for the limit of the function $v(\lambda,\lambda,\lambda)$. Comparing this to~\eqref{eq:tau}, we see the constants above get fixed and we obtain the following in the Kleinian limit:
\begin{equation}
a = i \pi, \quad b=c=0, \quad d=1 \implies \lim_{\lambda \to 0}  \rho_1(z) = e^{i \pi \tau(z)} =z_1.
\end{equation}
Note that we can repeat the same procedure for other punctures and similarly obtain $\lim_{\lambda \to 0}  \rho_j = z_j$ up to an unimportant phase factor. We explicitly checked this is indeed the case.

Now observe the scale factor $N=e^{-\frac{\pi}{2\lambda}}$ that relates the actual local coordinates $w_j$ to $\rho_j$ by $N w_j =\rho_j$ approaches to zero as $\lambda \to 0$, which is essentially a consequence of shrinking grafted cylinders. So in order to get a well-defined limit, it is necessary to place a \emph{cut-off} on the scale factor $N$ which we can do it as follows. We know $\lambda \to 0$ would make the length of the boundary geodesics $L$ smaller. So, as we take this limit, we choose some value of $L = \epsilon \ll 1$ that we put in $N$ and keep using it for any $L < \epsilon $. In other words, we take $N\approx e^{-\frac{\pi^2}{\epsilon }}$ for sufficiently small $L = 2 \pi \lambda \geq 0$ instead of what is given before. Note that this procedure essentially mirrors what is done in~\cite{Moosavian:2017qsp,Moosavian:2017sev}.

With this cut-off in place, we now have $N w_j = \rho_j = z_j$ in the Kleinian limit. Like in~\cite{Moosavian:2017qsp,Moosavian:2017sev}, we will multiply the original local coordinates $z_j$ for the Kleinian vertex given in (\ref{eq:KleinianVertex}) by $N^{-1}$ and define a new set of local coordinates $z_j' \equiv N^{-1} z_j$ in order to use the standard plumbing parameters. With this, we get $\lim_{\lambda \to 0} w_j = z_j'$ and see the scaled local coordinates for the Kleinian vertex matches what we find from the Kleinian limit of the hyperbolic three-string vertex as anticipated.

\subsection{Light-cone vertex}

Lastly, consider the situation
\begin{equation}
\lambda_1 = r \lambda, \qquad \lambda_2 = \lambda, \qquad \lambda_3 = (1-r) \lambda,
\end{equation}
for $0<r<1$ and take $\lambda \to \infty$. Having $\lambda_{2}=\lambda_{1}+\lambda_{3}$ while all of them being large, this limit should produce the local coordinates for the light-cone vertex~\cite{Zwiebach:1988qp}, by using similar geometric reasoning given in subsection \ref{sec:Witten}. Thus, we are going to call this limit the \emph{light-cone limit}.

In order to understand this limit better, first note that the restriction $\lambda_{2}=\lambda_{1}+\lambda_{3}$ always makes one of the first two arguments of the hypergeometric function appearing in local coordinates~\eqref{eq:lc} independent of $\lambda$ and finite as $\lambda \to \infty$. This is crucial because then a generic term in the expansion of these hypergeometric function around the puncture $z=z_j$ takes the following form:
\begin{equation}
\text{term} \sim \frac{\# \lambda^n + \dots }{\# \lambda^n + \dots} (z-z_j)^n.
\end{equation}
Here $\#$ denotes some numbers while dots denote the lower order terms in $\lambda$. The important point here is that since one of the arguments of the hypergeometric function is independent of $\lambda$, the same power of $\lambda$ appears in the numerator and the denominator of the coefficient of $(z-z_j)^n$ in its expansion. Therefore, these coefficients remain finite as we take $\lambda \to \infty$.

On the other hand, observe that the ratio of hypergeometric functions is raised to the power $1/i\lambda$ in~\eqref{eq:lc} and this exponent approaches to $0$ in the light-cone limit. But as we have just argued, the expansion of the hypergeometric functions remains finite in this limit. So we conclude that the part depending on the hypergeometric functions must completely drop out. The resulting limit gives, after taking the limits of prefactors like in subsection~\ref{sec:Witten},
\begin{subequations}
\begin{align}
w_1(z) &= r^{-1} (r-1)^{\frac{r-1}{r}}z (1-z)^{-\frac{\lambda_{2}}{\lambda_{1}}},\\ 
w_2(z) &= r^{r} (r-1)^{1-r} (z-1) z^{-\frac{\lambda_{1}}{\lambda_{2}}}, \\
w_3(z) &= r^{\frac{r}{r-1}} (r-1)^{-1} (z-1)^{-\frac{\lambda_{2}}{\lambda_{3}}} z^{\frac{\lambda_{1}}{\lambda_{3}}}.
\end{align}
\end{subequations}
From this, it is clear that if we relate the lengths of strings $2 \pi \lambda_j = L_j$ to the light-cone momenta $p_j^+$ in the usual fashion after an infinite rescaling, i.e. $p_j^+ \sim L_j$, and include the appropriate signs for the incoming/outgoing momenta, we arrive the light-cone vertex given in~\cite{Zwiebach:1988qp} up to an unimportant phase ambiguity. Therefore, the hyperbolic three-string vertex indeed reduces to the light-cone vertex in the light-cone limit in accord with our geometric expectation.

\section{Conservation laws for the hyperbolic three-string vertex} \label{sec:Conservation}

In this section we derive the conservation laws associated with the hyperbolic three-string vertex in the spirit of~\cite{rastelli2001tachyon}. Let us denote the hyperbolic three-string vertex with the geodesic boundaries of length $L=2 \pi \lambda$ as $\bra{V_{0,3}(\lambda)}$. This should be thought as an element of three-string dual Fock space, so it takes 3 states in Fock space and maps to a complex number. For simplicity of reporting, we set all the boundary lengths equal and report the holomorphic Virasoro conservation laws, but arguments here can be extended trivially to the cases with unequal lengths; ghosts and current conservation laws; and/or anti-holomorphic analogues.

First, let us put the punctures at $z=\sqrt{3},0, - \sqrt{3}$ in order to be consistent with~\cite{rastelli2001tachyon} and report the expansions for $z$ in terms of the local coordinates~$w_j$:
\begin{align}
f_1(w_1) &= \sqrt{3} + 2 \sqrt{3} N e^{-\frac{v}{\lambda }} w_1 + 3 \sqrt{3} N^2 e^{-\frac{2 v}{\lambda }} w_1^2 \nonumber
+\frac{\sqrt{3} \left(31 \lambda ^2+139\right)}{8 \left(\lambda ^2+4\right)}N^3 e^{-\frac{3 v}{\lambda }} w_1^3 + \mathcal{O}(w_1^4), \\
f_2(w_2) &= \frac{1}{2} \sqrt{3} N e^{-\frac{v}{\lambda }} w_2  -\frac{5 \sqrt{3} \left(\lambda ^2+1\right) N^3 e^{-\frac{3 v}{\lambda }}}{32 \left(\lambda ^2+4\right)} w_2^3 + \mathcal{O}(w_2^5),\nonumber\\
f_3(w_3) &= -\sqrt{3} + 2 \sqrt{3} N e^{-\frac{v}{\lambda }} w_3 - 3 \sqrt{3} N^2 e^{-\frac{2 v}{\lambda }} w_3^2
+\frac{\sqrt{3} \left(31 \lambda ^2+139\right)}{8 \left(\lambda ^2+4\right)} N^3 e^{-\frac{3 v}{\lambda }} w_3^3 + \mathcal{O}(w_3^4). 
\end{align}
These are based on inverting~\eqref{eq:lc} respectively after performing the global conformal transformation
\begin{equation} \label{eq:move}
z \to - \frac{z-\sqrt{3}}{z+\sqrt{3}},
\end{equation}
that makes the monodromies around $z=\sqrt{3},0, -\sqrt{3}$ non-trivial. Here we will refer functions from the $w_j$-plane to the $z$-plane as $f_j$, $f_j(w_j) = z$. Like before, here $N=e^{-\frac{\pi}{2 \lambda}}$ and $v=v_1=v_2=v_3$. The global phase of the local coordinates $w_j$ are not important as usual, so we used this freedom to put $f_j$'s into rather symmetric form shown above. We are going to work perturbatively in $w_i$ below.

Notice that when we consider the minimal area limit, these expressions reduce to the one given in (2.11) of~\cite{rastelli2001tachyon}. This limiting behavior is expected, since the vertex given there, open string Witten vertex, when considered in the entirety of the complex plane becomes the closed string minimal area three-vertex, and we know from the previous sections that's what the hyperbolic three-string vertex approaches in the minimal area limit. So it shouldn't be too surprising that the identities we will write below reduces to their counterparts given in~\cite{rastelli2001tachyon} in the minimal area limit.

As an example of conservation laws, we derive the Virasoro conservation laws by which we mean the identities of the type, for $k>0$,
\begin{equation} \label{eq:Form}
\bra{V_{0,3}(\lambda)} L_{-k}^{(2)}  = \bra{V_{0,3}(L)} \left[A^k(\lambda) \cdot c + \sum_{n\geq 0 } a_n^k(\lambda) L_n^{(1)}+ \sum_{n\geq 0 } c_n^k(\lambda) L_n^{(2)}+ \sum_{n\geq 0 } d_n^k(\lambda) L_n^{(3)}\right].
\end{equation}
Here $A^k, a_n^k, c_n^k, d_n^k$ are some functions of $\lambda$ that we are going to explicitly derive, $L_n$ are Virasoro generators, and the superscript denotes the slot that they apply in $\bra{V_{0,3}(\lambda)}$. By cyclicity of the hyperbolic three-vertex similar identities holds as we permute $(1) \to (2), (2) \to (3), (3) \to (1)$. So it would be sufficient to report the form above. The idea here is to exchange the negatively-moded Virasoro charges with the positively-moded ones plus the central term.

Now let $v(z)$ be a vector field holomorphic everywhere except for the punctures.\footnote{Not to be confused with $v$ appearing in the local coordinates.} That is, it changes as $v(z) \to \tilde{v}({\tilde{z}}) = (\partial \tilde{z}) v(z)$ under $z \to \tilde{z}$. Note that $v(z)$ should be regular at $z=\infty$ by its definition, so we must ensure $z^{-2}v(z)$ is finite as $z \to \infty$ by the inversion map $z \to \tilde{z} = 1/z$. 

It is important to note that the object $v(z) T(z) dz$ is almost a 1-form, where $T(z)$ is stress-energy tensor.\footnote{In this section $T(z)$ will denote the stress-energy tensor of an arbitrary CFT with central charge $c$, not to be confused with the stress-energy tensor $T_{\varphi}(z)$ we previously considered.} Under $z \to \tilde{z}$ it transforms as,
\begin{equation} \label{eq:formtransform}
v(z) T(z) dz = \tilde{T}(\tilde{z}) \tilde{v}(\tilde{z}) d \tilde{z} - \frac{c}{12} \{z, \tilde{z}\} \tilde{v}(\tilde{z}) d \tilde{z}.
\end{equation}
As we see above, we have an extra contribution from the central term. Nonetheless, we can integrate this object on the complex plane on contours and use the usual properties of the complex integration, as long as we keep track of this additional term under the change of coordinates.

In order to derive the Virasoro conservation laws, the following equality is crucial:
\begin{equation}
\bra{V_{0,3}(\lambda)} \oint_{\mathcal{C}} \text{d} z \; v(z) T(z) = 0.
\end{equation}
Here, $\mathcal{C}$ is a contour that surrounds the three punctures, oriented counterclockwise. This is a shorthand notation for the vanishing of the correlator of the integral $\oint_{\mathcal{C}} v(z) T(z) dz$ with any three vertex operator placed at the punctures. Note that this correlator vanishes because we can push the contour to shrink around $z=\infty$ by the inversion map. In this case, the central charge term does not contribute since the Schwarzian derivative of the inversion map is zero.

Now we can deform the contour $\mathcal{C}$ to separate it into positively oriented, disjoint contours $\mathcal{C}_i$ around each punctures and write down the expression above in terms of the local coordinates as follows:
\begin{equation} \label{eq:conservation}
\bra{V_{0,3}(\lambda)} \sum_{i=1}^{3} \oint_{\mathcal{C}_i} \text{d} w_i \; v^{(i)}(w_i) \left[T^{(i)}(w_i) - \frac{c}{12} \{f_i(w_i), w_i\}\right] = 0,
\end{equation}
using the transformation property of $v(z) T(z) dz$ given in~\eqref{eq:formtransform}. Here $v^{(i)}(w_i)$ denotes the components of the vector field $v(z)\frac{\partial}{\partial z}$ in the local coordinates $w_i$ and similarly for the stress-energy tensor.

We will clearly need to find $\{f_i(w_i), w_i\} $ because of~\eqref{eq:conservation}. This is easy to do:
\begin{align}
\{f_i(w_i), w_i\} &= -\frac{15 \left(\lambda ^2+1\right)}{8 \left(\lambda
	^2+4\right)} N^2 e^{-\frac{2 v}{\lambda }} + \frac{135 \left(\lambda ^2+1\right) \left(3 \lambda ^4+19 \lambda ^2-2\right)}{64 \left(\lambda ^2+4\right)^2 \left(\lambda ^2+16\right)} N^4  e^{-\frac{4 v}{\lambda }} w_i^2 + \cdots.
\end{align}
This is the same for each puncture because of the cyclicity, which we explicitly checked. Unsurprisingly, in minimal area limit we arrive the expression given in equation (3.8) of~\cite{rastelli2001tachyon}. By this expansion it is easy to see this term only appears if we have odd-powered poles around a puncture by (\ref{eq:conservation}).

Now remember
\begin{equation}
L_{-k}^{(i)} = \oint_{\mathcal{C}_i} \frac{\text{d} w_i}{2\pi i} w_i^{-k+1} T^{(i)}(w_i),
\end{equation}
so we need a vector field that behaves like $v^{(2)} \sim w_2^{-k+1}$ for $k>0$ around the puncture $(2)$ while behaves like $v^{(1)} \sim w_1$ and $v^{(3)} \sim w_3$ around the other punctures in order to put the Virasoro generators in the form given in (\ref{eq:Form}). Additionally, we have to ensure the regularity at infinity.

For $k=1$ case, all of these can be achieved with the following globally defined holomorphic vector field:
\begin{equation}
v_1(z) = -\frac{N e^{-\frac{v}{\lambda }}}{2 \sqrt{3}}  \left(z^2-3\right).
\end{equation}
Normalization is chosen to get the convention in~(\ref{eq:Form}) and in the minimal area limit this reduces to one given in (3.10) of~\cite{rastelli2001tachyon}. This has the following expansion in the local coordinates $w_i$
\begin{align}
v_1^{(1)}(w_1) &= -Ne^{-\frac{v}{\lambda }}w_1 + \frac{1}{2} N^2 e^{-\frac{2 v}{\lambda }} w_1^2 - \frac{5 \left(\lambda ^2+1\right) }{8 \left(\lambda 
	^2+4\right)} N^3 e^{-\frac{3 v}{\lambda }} w_1^3 + \mathcal{O}(w_1^4), \nonumber\\
v_1^{(2)}(w_2) &= 1 + \frac{\left(11 \lambda ^2-1\right)}{16 \left(\lambda
	^2+4\right)} N^2 e^{-\frac{2 v}{\lambda }}w_2^2 + \frac{5 \left(-8 \lambda ^6-6 \lambda ^4+3 \lambda ^2+1\right) }{128 \left(\lambda ^2+4\right)^2 \left(\lambda ^2+16\right)} N^4 e^{-\frac{4 v}{\lambda
}} w_2^4 + \mathcal{O}(w_2^6),\nonumber\\
v_1^{(3)}(w_3) &= Ne^{-\frac{v}{\lambda }}w_3 + \frac{1}{2} N^2 e^{-\frac{2 v}{\lambda }} w_3^2 + \frac{5 \left(\lambda ^2+1\right) }{8 \left(\lambda 
	^2+4\right)} N^3 e^{-\frac{3 v}{\lambda }} w_3^3 +  \mathcal{O}(w_1^4).
\end{align}
Unsurprisingly, these reduce to the equation (3.11) of \cite{rastelli2001tachyon} in the minimal area limit. After substituting this into (\ref{eq:conservation}), each integration amounts to doing the replacement $w_i^n \to L_{n-1}^{(i)}$ by the residue theorem. Therefore we get 
\small
\begin{alignat}{2}
0 &= \bra{V_{0,3}(\lambda)}  \left( -Ne^{-\frac{v}{\lambda }}L_0  + \frac{1}{2} N^2 e^{-\frac{2 v}{\lambda }} L_1 - \frac{5 \left(\lambda ^2+1\right) N^3 e^{-\frac{3 v}{\lambda }}}{8 \left(\lambda 
	^2+4\right)} L_2 + \frac{5 \left(\lambda ^2+1\right) N^4 e^{-\frac{4 v}{\lambda }}}{32 \left(\lambda ^2+4\right)}L_3 + \dots  \right)^{(1)}\nonumber\\
&+\bra{V_{0,3}(\lambda)} \left(L_{-1} + \frac{\left(11 \lambda ^2-1\right) N^2 e^{-\frac{2 v}{\lambda }}}{16 \left(\lambda
	^2+4\right)} L_1 + \frac{5 \left(-8 \lambda ^6-6 \lambda ^4+3 \lambda ^2+1\right) N^4 e^{-\frac{4 v}{\lambda
}}}{128 \left(\lambda ^2+4\right)^2 \left(\lambda ^2+16\right)} L_3 + \dots \right)^{(2)}\nonumber\\
&+\bra{V_{0,3}(\lambda)} \left( Ne^{-\frac{v}{\lambda }}L_0  + \frac{1}{2} N^2 e^{-\frac{2 v}{\lambda }} L_1 + \frac{5 \left(\lambda ^2+1\right) N^3 e^{-\frac{3 v}{\lambda }}}{8 \left(\lambda 
	^2+4\right)} L_2+ \frac{5 \left(\lambda ^2+1\right) N^4 e^{-\frac{4 v}{\lambda }}}{32 \left(\lambda ^2+4\right)}L_3 + \dots  \right)^{(3)}.
\end{alignat}
\normalsize
Again, this reduces to (3.12) of~\cite{rastelli2001tachyon} in the minimal area limit. Note that this doesn't have any central charge contribution since the vector $v_1(z)$ does not have a pole around the punctures.

We can continue to generate identities of the form (\ref{eq:Form}) by using the following vector fields:
\begin{align}
v_2(z) &= -\frac{N^2 e^{-\frac{2v}{\lambda }}}{4}  \frac{z^2-3}{z},\\
v_3(z) &= -\frac{\sqrt{3 }N^3 e^{-\frac{3v}{\lambda }}}{8}  \frac{z^2-3}{z^2}-
\frac{3 (3+7\lambda^2) }{16(4+\lambda^2)} N^2 e^{-\frac{2v}{\lambda }} v_1(z).
\end{align}
They produce the following identities respectively,
\small
\begin{alignat}{2}
0 &= \bra{V_{0,3}(\lambda)}  \left( -\frac{1}{2} N^2 e^{-\frac{2 v}{\lambda }} L_0
+\frac{5}{4} N^3 e^{-\frac{3 v}{\lambda }} L_1
-\frac{3 \left(7 \lambda ^2+23\right) }{16 \left(\lambda
	^2+4\right)} N^4 e^{-\frac{4 v}{\lambda }} L_2  + \dots \right)^{(1)}\nonumber\\
&+\bra{V_{0,3}(\lambda)} \left( L_{-2} + \frac{5 \left(\lambda ^2+1\right) }{32 \left(\lambda
	^2+4\right)} N^2 e^{-\frac{2 v}{\lambda }} c + \frac{\left(4 \lambda ^2+1\right)}{4 \left(\lambda
	^2+4\right)} N^2 e^{-\frac{2 v}{\lambda }} L_0 + \dots \right)^{(2)}\nonumber\\
&+\bra{V_{0,3}(\lambda)}  \left( -\frac{1}{2} N^2 e^{-\frac{2 v}{\lambda }} L_0
-\frac{5}{4} N^3 e^{-\frac{3 v}{\lambda }} L_1
-\frac{3 \left(7 \lambda ^2+23\right)}{16 \left(\lambda
	^2+4\right)} N^4 e^{-\frac{4 v}{\lambda }} L_2+ \dots \right)^{(3)}, \\
0 &= \bra{V_{0,3}(\lambda)}  \left(\frac{\left(17 \lambda ^2-7\right) }{16 \left(\lambda
	^2+4\right)} N^3 e^{-\frac{3 v}{\lambda }} L_0 + \frac{15 \left(\lambda ^2+9\right) }{32 \left(\lambda
	^2+4\right)} N^4 e^{-\frac{4 v}{\lambda }} L_1 + \dots \right)^{(1)}\nonumber\\
&+\bra{V_{0,3}(\lambda)} \left(L_{-3} -\frac{15 \left(\lambda ^2+9\right) \left(2 \lambda ^2-1\right) \left(4 \lambda ^2+1\right) }{128 \left(\lambda ^2+4\right)^2 \left(\lambda ^2+16\right)} N^4 e^{-\frac{4 v}{\lambda
}} L_1 + \dots \right)^{(2)} \nonumber \\
&+\bra{V_{0,3}(\lambda)}  \left(-\frac{\left(17 \lambda ^2-7\right) }{16 \left(\lambda
	^2+4\right)} N^3 e^{-\frac{3 v}{\lambda }}L_0 + \frac{15 \left(\lambda ^2+9\right) }{32 \left(\lambda
	^2+4\right)} N^4 e^{-\frac{4 v}{\lambda }} L_1 + \dots \right)^{(3)}. 
\end{alignat}
\normalsize
We explicitly checked these reduces to their counterparts in~\cite{rastelli2001tachyon} in the minimal area limit. Note that we can continue generating similar identities for $L_{-k}$ recursively by using vector fields $v_k(z) \sim (z^2-3)z^{-k+1}$ and appropriately subtracting previous ones. Doing this allows us to put the identities in the form (\ref{eq:Form}) for which only a single negatively-moded Virasoro generator appears in the left-hand side.

\section{Remarks and open questions} \label{sec:Conc}

In this paper, we constructed the local coordinates for the hyperbolic three-string vertex first described in~\cite{Costello:2019fuh} and investigated its various limits explicitly. We calculated the $t^3$ term in the closed string tachyon potential and developed the conservation laws associated with such vertex in the spirit of~\cite{rastelli2001tachyon}. We conclude by providing some final remarks and highlighting possible future directions relevant to us:
\begin{enumerate}
	\item Since we now know the local coordinates for the hyperbolic three-string vertex, it is possible to construct the Feynman diagrams by identifying them as
	\begin{equation}
	w_j w_j' = \exp \left[ -\frac{2 \pi s}{L_j} + i \theta\right] \quad \text{with} \quad s \in \mathbb{R}_{\geq 0}, \quad \theta \in [0,2 \pi)
	\end{equation}
	using the local coordinates $w_j$ and $w_j'$ associated to boundaries of equal length on not-necessarily-distinct pair of pants. Making this identification corresponds to having a finite flat cylinder of circumference $L_j$ and length $s$ with a twist $\theta$ stretching between not-necessarily-distinct pair of pants and it has the natural interpretation of the string propagator. 
	
	As usual, we must consider every possible value of $(s,\theta)_A$ when we are computing the string amplitudes. Here we added an index $A$ to indicate there are generally more than one propagator in the Feynman diagrams. It would be interesting the study the Feynman regions these diagrams cover in the moduli spaces of Riemann surfaces of genus $g$ and $n$ punctures $\mathcal{M}_{g,n}$ to see if they provide a piece of a section over the bundle $\widehat{\mathcal{P}}_{g,n} \to \mathcal{M}_{g,n}$ or not. The simplest Feynman regions to study would be for four-string scattering or string tadpole interaction. Note that with the metric we constructed on the hyperbolic pair of pants, it is possible to describe a Thurston metric of~\cite{Costello:2019fuh} explicitly on Riemann surfaces.
	
	\item The local coordinates (\ref{eq:LocalCoord}) we constructed in this paper also can be used for the open-closed hyperbolic string vertices without moduli~\cite{Cho:2019anu}. There are two additional vertices without moduli on top of the sphere with three closed string punctures in this situation. They are disk with three open string punctures and disk with one open string puncture and one closed string puncture. Note that if we cut open the hyperbolic three-closed string vertex along a geodesics connecting all punctures for the former and one puncture connecting back to itself for the latter, we generate these additional cases exactly. From this construction it is clear that these would carry hyperbolic metric with appropriately grafted flat strip/cylinder parts and would be the same as what is constructed in~\cite{Cho:2019anu}. So we can still use the local coordinates (\ref{eq:LocalCoord}) for these additional cases.
	
	\item The primary method we used in this paper, that is relating Liouville's equation on a specified domain to a monodromy problem, can be generalized to construct the local coordinates of the classical ($g=0$) hyperbolic $n$-string vertices in principle.  For this, instead of (\ref{eq:3T}) we should take the stress-energy tensor of Liouville's equation to be
	\begin{equation}  \label{eq:Tconc}
	T_{\varphi}(z) = \sum_{i=1}^n \left[ \frac{\Delta_i}{(z-z_i)^2} + \frac{c_i}{(z-z_i)}\right] ,
	\end{equation}
	with punctures positioned at $z=z_i$ and use this stress-energy tensor to generate the local coordinates. Here $c_i \in \mathbb{C}$ are so-called \emph{accessory parameters}~\cite{hadasz2003polyakov}.
	
	There are two important problems with this approach. First, after we fixed the positions of three punctures by PSL(2,$\mathbb{C}$) symmetry, assigned prescribed weights at all punctures, and demanded regularity at infinity, we would still have $n-3$ unfixed $c_i$ parameters functions of $n-3$ unfixed positions $z_i$, the usual moduli for the $n$-punctured sphere. It is argued that such accessory parameters can be fixed in terms of moduli using the action of Liouville theory so that the metric associated to $T_{\varphi}(z)$ is smooth and hyperbolic~\cite{hadasz2003polyakov}, which goes under the name \emph{Polyakov Conjecture}. Computing these parameters exactly is not known, so this is the first problem. However, some numerical results are available in the case of vanishing $L_i$, see~\cite{hadasz2006liouville}.
	
	Secondly, even if we find the correct $c_i$, guaranteeing the correct monodromy structure for the resulting Fuchsian equation with $n$ regular singularities is impractical. This is because of the lack of analogous formulas given in (\ref{eq:connection}) for the solutions to the Fuchsian equation associated with~\eqref{eq:Tconc}. It seems to us this is not the direction one should pursue if their goal is to do practical computations.
	
	\item It would seem more promising to evaluate all higher elementary string interactions by exploiting the pants decomposition of the (marked) Riemann surfaces and their associated Teichm\"uller spaces~\cite{buser2010geometry}. The idea would be to decompose the contribution from a given Riemann surface as sums of products of cubic interaction of appropriate string fields dictated by a given pair of pants decomposition of such Riemann surface and to use an appropriate region in Teichm\"uller space to perform the moduli integration, similar to what is suggested in~\cite{Moosavian:2017qsp,Moosavian:2017sev}. Both of these steps need further study. Related to this idea, it may be possible to form a recursion relations in the similar vein of~\cite{mirzakhani2007weil,Eynard:2007fi,ellegard1,ellegard2}.
	
	From the possibility of using pants decomposition we see the relevance of the hyperbolic three-string vertex with \emph{unequal} $L_i$ we considered so far. After the pants decomposition, we would only need to use the hyperbolic three-string vertex of arbitrary $L_i$ to compute CFT correlators and the rest of the computation would presumably just involve combining them together in correct fashion. We leave investigating this to a future work.
\end{enumerate}

\acknowledgments

The author would like to thank Barton Zwiebach for suggesting this problem and his guidance in the writing process. This material is based upon work supported by the U.S. Department of Energy, Office of Science, Office of High Energy Physics of U.S. Department of Energy under grant Contract Number  DE-SC0012567.


\end{document}